\documentclass[12pt,preprint]{aastex}

\def\kms{\relax \ifmmode {\ \rm km s}^{-1}\else \ km\ s$^{-1}$\fi}
\def\ergcms{${\rm erg}^{-1}\ {\rm cm}^{-2}\ {\rm s}^{-1}$} 

\def\Mso{{$M_{\rm \odot}$}}

\def\cm3{${\rm cm}^{-3}~$}

\def\nii{[N~{\sc ii}]}

\def\heii{He~{\sc ii}}
\def\oiii{[O~{\sc iii}]}
\def\ha{H$\alpha~$}

\slugcomment{}

\shorttitle{The formation and evolution of PNe}
\shortauthors{Villaver, Manchado, \& Garc\'{\i}a-Segura}
  
\begin{document}
  
\title{The Dynamical Evolution of the Circumstellar Gas around Low- and
 Intermediate-Mass stars.II. The Planetary Nebula formation.}
  
\author{Eva Villaver\altaffilmark{1,2}, Arturo Manchado\altaffilmark{1,3}, \& 
Guillermo Garc\'{\i}a-Segura\altaffilmark{4}}

\altaffiltext{1}{Instituto de Astrof\'{\i}sica de Canarias, V\'{\i}a
        L\'actea 
        S/N, E-38205 La Laguna, Tenerife, Spain
{\tt villaver@ll.iac.es, amt@ll.iac.es}} 
\altaffiltext{2}{Space Telescope Science Institute, 3700 San
        Martin Drive, Baltimore, MD 21218; {\tt villaver@stsci.edu}} 
\altaffiltext{3}{Consejo Superior de Investigaciones Cientificas, Spain}
\altaffiltext{4}{Instituto de Astronom\'{\i}a-UNAM, Apartado postal 877,
       Ensenada, 22800 Baja California, M\'exico
 {\tt ggs@astrosen.unam.mx}}

\begin{abstract}
We have studied the effect of the mass of the central star (CS) on the
gas evolution during the planetary nebula (PN) phase. We have performed
numerical simulations of PN formation using CS tracks 
for six stellar core masses corresponding to initial masses from 1 to 5
\Mso. The gas structure resulting from 
the previous asymptotic giant branch (AGB) evolution is used as the starting
configuration. The formation of multiple shells is discussed in the light of
our models, and 
the density, velocity and \ha emission brightness profiles are shown for each
stellar mass considered. We have computed the evolution of the different
shells in 
terms of radius, expansion velocity, and \ha peak emissivity. We find
that the evolution of the main shell is controlled by the ionization
front rather than by the thermal pressure provided by the 
hot bubble during the
early PN stages. This effect 
explains why the kinematical ages overestimate the
age in young CSs. At later stages in the
evolution and for low mass progenitors the kinematical ages severely
underestimate the CS age. Large (up to 2.3 {\rm pc}), low surface
brightness shells (less than 2000 times the brightness of the main
shell) are formed in all of our models (with the exception of the 5~\Mso\ 
model). These PN halos contain most of the ionized mass in PNe, which we find
is greatly underestimated by the observations because of the low surface
brightness of the halos. 
\end{abstract}


\section{INTRODUCTION}
Low-and intermediate-mass stars
experience high mass loss rates during the asymptotic giant branch
(AGB) phase. Owing to the high mass loss rates during this phase, 
stars which have masses between 1 and $\sim$8 \Mso~end their lives as
white dwarfs with final masses below the Chandrasekhar 
mass limit (1.44~\Mso). 

After an uncertain phase of evolution called the transition time (the
timescales are not completely known), the stellar
remnant\footnote{Central 
star (CS) from now onwards.} becomes hot enough
to ionize the previously ejected envelope. In the meantime, the wind
velocity increases and shapes the inner parts of the envelope according to
the so-called interacting stellar winds \citep{Kpf:78}.
This leads to the formation of a planetary nebula (PN). The evolution of the
nebular gas from then onward depends on the         
energy provided by the CS through the 
wind and the radiation field, and both depend on the stellar luminosity and
effective temperature. The post-AGB evolution of the CS is mainly determined
by its core mass 
(\citealt{Pac:71}; \citealt{Wf:86}; \citealt{Vw:94} [hereafter VW94]) and by
the previous AGB evolution \citep{Blo:95}. 

The relationship between the CS
evolution and that of the nebular shells has been the subject of many
observational studies 
(e.g., \citealt{Scs:93}; \citealt{Fvb:94}; \citealt{Sp:95};
\citealt{Getal:96}; 
\citealt{Hetal:97};\citealt{Setal:02}). However, most of 
the numerical studies of PN evolution in the literature have been restricted
to a $\sim$0.6~\Mso~post-AGB evolutionary track, without allowing
for the CS mass range. This lack of consideration for the CS mass
range is not the 
only problem with the existing numerical models of PN formation. Classically,
the complex AGB evolution have 
been modeled by a simple $r^{-2}$ density law \citep{Oetal:85,Sk:87,Fbr:90,
Ms:91,Mel:94}. In the series of related papers of \cite{Setal:97},
\cite{Petal:98}, and \cite{Cetal:00}, the previous 
AGB history is included; however, the grids were truncated to
$\sim$0.65 pc, and hence 
only the recent mass loss history can be completely followed up. Thus, the
consequences of the 
long term thermal-pulse AGB evolution on PN structure at large
scales still need to be tested.  

Our aim in this paper is to study the role that the stellar evolution and the
stellar progenitor mass play on PN formation.
In a previous paper (\citealt{Vgm:02} [hereafter, Paper I]) we described the
dynamical 
evolution of the stellar wind during the AGB for low- and intermediate-mass
stars. In this paper we
use the gas structure resulting from the previous AGB evolution as the
starting configuration. The PN formation is then followed by using post-AGB
tracks that comprise the mass range of 0.57 to 0.9~\Mso, which correspond to
main sequence masses of between 1 and 5~\Mso \citep{Vw:93}. In $\S2$ we
describe the numerical 
method. The results of our models for 
different PN progenitors and grid sizes are presented in Section 3. In $\S4$
we 
describe  
the observational properties derived from our models and we compare them with
observations. Our conclusions are summarized in Section~5.

\section{NUMERICAL METHOD AND COMPUTATIONAL DETAILS}
The numerical simulations were done with 
the multi-purpose fluid solver ZEUS-3D (version 3.4), developed by
M.~L. Norman and the Laboratory for Computational Astrophysics. This is a 
finite-difference, Eulerian, fully explicit code with a modular structure
that allows to solve a wide range of astrophysical problems (for further 
details of the numerical method see \citealt{Sn:92a,Sn:92b,Smn:92}).

By choosing two symmetry axis we perform 1D simulations
in spherical coordinates. We use
two grid sizes for each progenitor mass considered; a small one to study the
development of structures in the 
inner parts of the nebula (between 0.5 and 1 {\rm pc}), and a large
one (out to 3 {\rm pc}) to study the large-scale structures arising as a
consequence of the mass loss during the AGB phase. All the models have a
resolution of 1000 zones in the radial coordinate. 

The wind evolution is implemented by setting the velocity, mass loss
rate, and wind temperature in the five innermost 
zones of the grid. A free-streaming
boundary condition was set in the outer portion of the grid. As the CS
evolves, the number of ionizing photons changes, 
therefore we consider the evolution of the radiation field from the star
together with the evolution of the stellar wind. 

Photoionization has important effects on
the dynamical evolution of the nebular gas. It
increases the temperature and the number 
of particles in the gas, and thus the pressure of the photoionized gas is
initially a   
factor of two hundred larger than the pressure of the surrounding neutral
material. 
The dynamical effects of this increase in pressure are not
negligible and must be taken into account. Since ZEUS-3D does not include
radiation transfer, we use the approximation 
implemented by \cite{Gf:96} to derive the location of 
the ionization front (IF) for arbitrary density distributions
(\citealt{Bty:79}; \citealt{Fttb:89}, 1990).  
The position of the IF can be determined by assuming that 
ionization 
equilibrium holds at all times, and that the
ionization is complete within the ionized
sphere and zero outside. This implies that we are
adopting the classical definition of the Str\"omgren sphere, where 
the position of the IF is given by
$\int n^2(r) r^2 dr \approx  Q_o/4 \pi \alpha_B $, where n(r) is the radial
density distribution, $Q_o$ is the number of ionizing photons, and $\alpha_B$ 
is the recombination coefficient to all excited levels. We apply this 
formulation by assuming that the
nebula is pure hydrogen, that it is
optically thick in the Lyman continuum, and that the `on the spot'
approximation is valid. However, we use solar composition for the
radiative cooling which is made according to the cooling curves of
\cite{Rs:77}, and \cite{Dm:72} for gas temperatures above $10^4$ {\rm K}
and according to \cite{Mb:81} for temperatures between $10^2$ and $10^4$ {\rm
  K}. The unperturbed gas is treated adiabatically. 
Finally, the photoionized gas is always kept at $10^4$ {\rm K}, so no cooling
curve is applied inside the photoionized region unless there is a shock.

\subsection{Initial Conditions}
We have set the zero-age for PN formation to be the time at which the star's
effective temperature is 10~000 {K}. The initial condition for each of our
models is the gas structure from the 
previous AGB evolution (see Paper I). In this paper we
assume that the transition time (the time elapsed between  AGB quenching
and the CS temperature reaching 10~000 {\rm K}) is zero. \cite{Sr:00} obtain
zero transition times when the stellar remnant evolves on a thermal
timescale. 

We have studied the evolution during the PN stage for six masses
whose CSs are modeled according to evolutionary tracks, taken from
VW94, corresponding to
stellar cores with initial masses of 1, 1.5, 2, 2.5, 3.5, and 5 
M$_{\rm \odot}$. In Figure~1 we show the density and velocity structure of
the gas 
used as the starting configuration for each model. In this plot we show the
gas structure at the end of the AGB when the evolution of the stellar wind
during the thermal-pulsing AGB phase is considered and the ISM
density is 1 \cm3 (see Paper I for further details).  

\subsection{Inner Boundary Conditions: CS Evolution}
The evolution of the stellar wind is computed by using the post-AGB
evolutionary sequences given by VW94 for hydrogen burners with solar
metallicity. This means that the stars leave the  
AGB while burning hydrogen in a shell. The initial and core masses of the
models are given in columns 1 and 2 of Table 1.

The mechanism that drives the winds of CSs (with velocities several
orders of magnitude higher than that experienced during the AGB phase) is the
transfer of photon   
momentum to the gas through absorption by strong resonance
lines \citep{Petal:88}. The dependence of mass loss rate and velocity 
with the luminosity ($L$) and effective temperature of the star ($T_{\rm
  eff}$) has been adopted from VW94. 

In order to find the dependence of the terminal wind velocity (v$_{\infty}$; 
the velocity of the wind far 
away from the CS) on $L$ and $T_{\rm eff}$, VW94 fit
a relation between $v_{\infty}/v_{\rm esc}$ and $T_{\rm eff}$ using data
compiled by 
\cite{Petal:88} for CSs of PNe. Using $v_{\rm esc}=\sqrt{2GM/R}$ as the
stellar surface escape velocity and Stefan's law, $v_{\infty}$ can be written
in the form, $v_{\infty}=\beta~M^{\frac {1}{2}} L^{-\frac {1}{4}} T_{\rm
  eff}^{1.52}$, where $\beta=16\times\ 10^{-8}G^2\pi\sigma$. 

The mass loss during the PN regime was obtained by 
VW94 by fitting the theoretical results of \cite{Petal:88} to determine the
slope of 
the relation between $\log  {\dot{M}}/{\dot{M}_{\rm lim}}$ and 
$T_{\rm eff}$ which combined with \.M$_{\rm lim}= L/cv_{\infty}$ (the maximum
mass loss rate obtained 
by assuming that all the momentum of the  
stellar radiation field is injected into the wind), and with the relation
obtained for the terminal wind velocity, gives: 
$\dot{M}=\alpha~M^{\frac{-1}{2}} L^{\frac{5}{4}} T_{\rm eff}^{-0.85}$, 
where $\alpha=16\pi\times 10^{11.92}\times c^{-1}G^2\sigma$. 

In Figure~2 we show the temporal dependence of the 
mass loss rate and wind velocity used for our
models. Note that for higher masses,
the CS evolution is faster, the wind velocities are higher, and the mass loss
rates are lower (different 
timescales are used in these plots for low- and intermediate-mass
progenitors). The normalized wind kinetic  
energy and momentum injection rates of the wind as a function of time are 
shown in Figure~3. Note that the shorter 
timescales of the evolution are for the higher main sequence mass.
The normalization factors for the wind momentum and
kinetic energy are given in columns 3 and 4 of Table 1.   

\begin{deluxetable}{cccc}
\tablenum{1}
\tablewidth{25pc}
\tablecaption{Model masses and wind normalization parameters}
\tablehead{\multicolumn{1}{c}{Initial mass}& 
\multicolumn{1}{c}{Core mass} & 
\multicolumn{1}{c}{Momentum} &
\multicolumn{1}{c}{Kinetic energy}\\
\multicolumn{1}{c}{[\Mso]}&
\multicolumn{1}{c}{[\Mso]}& 
\multicolumn{1}{c}{[{g~cm~s$^{-2}$}]}&
\multicolumn{1}{c}{[{\rm erg~s$^{-1}$}]}}
\startdata
1   & 0.569 & 9.3$\times 10^{25}$ & 2.1$\times10^{34}$ \\
1.5 & 0.597 & 1.5$\times10^{26}$ & 4.3$\times10^{34}$ \\
2   & 0.633 & 2.5$\times10^{26}$ & 9.1$\times10^{34}$  \\
2.5 & 0.677 & 3.4$\times10^{26}$ & 1.5$\times10^{35}$ \\
3.5 & 0.754 & 5.9$\times10^{26}$ & 3.6$\times10^{35}$ \\
5   & 0.9   & 1.4$\times10^{27}$ & 1.6$\times10^{36}$ \\
\enddata
\end{deluxetable}

\subsubsection{Ionizing Radiation from the Star}
The number of ionizing photons for hydrogen,
$Q_o$, is computed in terms of the position of the star in the 
Hertzsprung--Russell (HR) diagram
assuming the star radiates as a blackbody. The hydrogen
ionization radiation is shown in Figure~4 for the first 10~000 {\rm yr} of
the CS evolution. The rapid decrease in the
number of ionizing photons in Figure~4 for stars with initial masses $\leq$
2 \Mso~is caused by the sudden fall in the stellar luminosity when the
stellar models run out of nuclear fuel. This occurs earlier for higher mass
stars since the higher the mass, the faster the evolution of the CS in the HR
diagram is.  

\section{RESULTS}
\subsection{PN evolution on Small Scales} 
In order to allow a direct
comparison of our models with observations we have computed the \ha emission
brightness 
profiles (EBPs). The emission coefficient in the \ha line 
for a nebula with a temperature of 10~000~{\rm K} has the
form \citep{Ost:89} 
$4\pi j_{\rm H\alpha}=3.5 \times 10^{-25}\cdot N_{\rm e}\cdot N_{\rm p}$ 
in \ergcms. The
intensity emitted by each element of volume ($I_{\rm H\alpha}= 4 \pi
j_{H\alpha}$) in the \ha line is integrated for each position in the
nebula along the line of sight to derive the EBPs. 

In the following, we use the brightness profile to distinguish
between detached halos and attached shells. As defined in \cite{Sp:95},
detached halos have round, detached limb-brightened outer shells, while the
brightness profile of attached shell shows a change in slope rather than a dip
between the shells. The brightness radial dependence for the detached
halos has a form  between $r^{-2.4}$ and $r^{-5.0}$, while for the attached
shells it is almost linear \citep{Gvm:98}.  

The results of our simulations are shown in Figures 5--10. The evolution
of the density, 
velocity and the normalized \ha brightness radial
distributions are shown for each mass at times which are representative of
the gas evolution. 
These models were computed at radial scales that allow us a high resolution
study of the processes taking place in the inner parts of
the nebula. The position of the IF is 
marked in the density profiles as a dotted line and the
times at which the plots has been selected are indicated in the EBP panels.
Note that both the time at which the plots have been selected and the
time interval between them are different for each stellar mass.

As the CS evolves, the wind velocity increases reaching a velocity above
which the shocked gas 
is not able to cool down radiatively. An adiabatic
shock then develops at the interaction region between the high velocity
stellar wind and the dense material ejected previously during the
AGB. The thermal energy 
provided by the hot bubble formed by this adiabatic shock will subsequently
compress the gas in the innermost regions, forming a shell.   
In our models during the early phases of the evolution, either the hot bubble
is not formed, or it cannot compete with the increase in pressure caused by
the photoionization. 

In Figure 5 we show the evolution of the gas for the 1 \Mso~model. Owing to
the highly supersonic expansion velocity of the ionized region with respect to
the surrounding neutral material, a shock front is formed just ahead of the
IF. A thin shell can be seen in the density profiles (see the first three
left panels of Fig.~5) formed by the compression associated with the shock
caused by the IF. Once
the innermost denser regions have been ionized, the propagation of the
IF is very fast. The 
propagation velocity of the Str\"omgren radius depends
mainly on the ionization flux it receives from the star 
and on the density of the neutral gas. The
density of the neutral material decreases with radius outside the innermost
high density region, and in the early stages of the CS evolution the number
of ionizing photons increases with time (see Fig.~4). The pressure
gradients generated by the IF 
act whenever 
there exists a relative increase in density. Thus, the ionized gas will expand
until it reaches pressure balance with its surroundings, and so it can be
seen how the gas behind the IF expands towards the
position of the CS (see the first three panels of Fig.~5 and note the negative
gas velocities inward of the position of the density maximum). At this 
time, the hot bubble either is not yet formed or cannot provide enough
pressure to stall the inwards expansion. However, as the star evolves towards 
higher effective temperatures, the wind velocity increases and 
the hot bubble is formed.

At $\sim$4500 {\rm yr} the hot bubble is already shaping the
innermost layer of the gas (the compression associated with the hot bubble 
makes the density of this innermost shell comparable to the
one caused by the IF). At this time the EBP shows a double-peak in
the innermost parts of the nebula. The innermost peak is caused by    
the hot bubble; the outermost peak is the consequence of an inward shock wave 
caused by the photoionized gas. The latter moves towards the position of the
CS, and as the contact discontinuity 
advances, the two shells merge. At this time all the gas is
ionized and two attached shells that surround the innermost double peak
structure are also visible, one at $\sim$0.1 {\rm pc} and a fainter one 
at $\sim$0.15 {\rm pc}. These attached shells are not the consequence
of the passage of the IF, rather they are formed by the expansion of the 
H~{\scshape ii} region in a stratified medium. The intermediate attached
shell eventually becomes a detached shell with approximately 11$\%$ of the
brightness of the main inner shell and its expansion velocity is $\sim$22
\kms (see the bottom panel in Fig.~5).

In Figure 6 we show the evolution of the 1.5 \Mso~model. In this case we can
see that, as in the 1 \Mso~case the shell formed by the propagation of the
IF is the brightest one during the first 3000 {\rm yr} of evolution. At
5500 {\rm yr} this shell has become detached
and surrounds the shell formed in the innermost regions by the hot
bubble. This 
detached shell is still visible in the EBP panels at 8000 {\rm yr} and
propagates outwards with a velocity of about 30 \kms.

The evolution of the 2 \Mso~model is shown in Figure 7. The gas evolution and
the kinds of shells that are formed are very similar to those formed for the
1.5 \Mso~case, in spite of the difference in timescales, energy input, and
initial gas density 
conditions. The shell formed by the IF is again the brightest one during
the early stages of the evolution ($\sim$3000 yr) and it evolves into a
detached shell when the shell 
formed by the hot bubble appears in the innermost regions of the gas. 

The models with initial masses of 2.5, 3.5, and 5 \Mso~spend a very short time
on the horizontal, constant luminosity part of the HR diagram. Moreover, 
the injection of kinetic energy reaches its maximum value earlier for higher
mass progenitors (see Figs~2 and 3). In Figure 8 we present the evolution of
the 2.5 \Mso~model and we can see that it is not yet fully ionized after
5500 {\rm yr}, so that only one shell appears in the EBPs. This
single shell is associated with the propagation of the IF, since the
compression associated with the wind--wind interaction process has not yet
started. 

The evolution for the 3.5 \Mso~model 
is shown in Figure 9. Two shells appear in the nebular structure
when all the gas is ionized; a bright main shell and surrounding it 
an attached shell. The dynamical
evolution of the brigthest main shell is controlled by the kinetic energy
provided by 
the wind since the very early stages of the evolution. Note that at about
1000 {\rm yr} the wind kinetic energy has 
already reached its maximum value (see Fig.~3), and that, therefore, the hot
bubble is already efficiently compressing  the innermost layers of the gas.

Only one shell is formed in the model with 5
M$_{\rm \odot}$ (shown in Fig.~10) which is associated with the IF. The
luminosity drop at the turn-around point for this model
$\sim$150 {\rm yr} after the star has reached  
a temperature of 10000 {\rm K}. Note in Figs~3 and 4
the rapid decrease in the number of ionizing photons, wind momentum and
energy for this model. The EBPs are uniform with an enhancement at the edge. 
This is the only model that does not produce a multiple shell structure and 
the only one that results is an ionization-bounded PN.

Our models may be overestimating the pressure in the boundary
between neutral and ionized gas because of the approximation we used
for the photoionization.  
First, near the Str\"omgren radius 
the ionization degree decreases sharply over a distance which is
proportional to the mean free path of a photon in the neutral
gas. However, because of the mean free path photon dependency with 
density, as the density decreases the mean free path increases and
the edge between ionized and neutral material begins to have a 
non-negligible thickness with respect to the size of the Str\"omgren radius.
We have used the classical Str\"omgren sphere approximation for
the photoionization; therefore, the boundary
between the ionized and neutral material is a discontinuity in our
models that could lead to an overestimate in the pressure at the IF when the
density of the medium is very 
low. None the less, in our simulations the IF is never trapped in
the low density media and we do not expect  our gas velocities to be
overestimated owing to this effect. Second, we assumed that ionization
equilibrium holds all the time, and 
that ionization is complete inside the Str\"omgren radius. If the 
IF moves slowly (in case of very high density media
or low ionizing photon flux), recombination occurs within the ionized 
volume. The presence of neutral atoms inside the Str\"omgren volume
reduces the number of photons arriving at the IF, which therefore slows down
or even stalls.  This might  affect our    2.5 and 5
\Mso~models since these are the only ones for which the IF is trapped in 
the high density media for a long enough time for recombination to take
place.  A full consideration of radiation
transfer will not change our main results (for details of how radiation
transfer shapes PNe 
the reader is referred to \citealt{Mel:94, Ost:89, Cetal:00}). 

We can only compare the 2.5 \Mso~model computed on small scales since this is
the only model 
with a PN track close enough to the 0.6 \Mso~track considered in previous
numerical models available in the literature. We find good agreement
with the kinds of structures formed in the models of Mellema  (1994, performed
at 0.25 {\rm pc}) or Corradi et al. (2000, performed at 0.65 {\rm pc}) on
intermediate scales, despite the differences in the departure conditions,
stellar evolutionary models, and radiation transfer approach.
Intermediate detached and attached shells are also formed
in those models.

\subsection{PN Evolution on Large Scales}
We have shown in Paper I that large-scale structures appear in the
circumstellar gas when the evolution during the AGB phase is
considered. The shells formed during the AGB phase have not diluted by the
time the PN phase starts (see Fig.~1). With the aim of 
studying these outer shells during the PN phase we performed 
numerical simulations using grid sizes that ensure that we do not lose
matter at the outer boundary. Here we present these models performed on
larger radial scales as radial distributions of 
density, velocity, and \ha EBP diagrams in
Figs 11, 12, 13, 14, 15, and 16 for the range of
initial-mass models considered previously. In order to show on the same graph
the normalized emission of the brightest shell and the faint emission of the
outer shell, we have plotted the EBPs at two intensity scales.

The model with 1 \Mso~is shown in Figure 11. At 4000 {\rm yr} an outer
detached shell
with a radius of $\sim$1.3 {\rm pc} and 5000 times fainter than the
brightest inner shell\footnote{We will refer to this shell as the main shell
  from now onwards.} is present. The outer shell is formed as a consequence
of  
the density structure generated by the previous AGB evolution. It is not a
signature of a thermal pulse and is formed owing to a shock in the interaction
region between two consecutive enhancements of the mass loss rate during the
AGB (see Paper I). A similar detached shell is present in the
other models (Figs~12,
13, 14, and 15), but in the one with 5 \Mso~(Fig.~16). The radius
of the outer shell in the 1.5, 2, 2.5, and 3.5 \Mso~models is $\sim$2 {\rm
  pc}. The ratio between  
the emission of the main shell and that of
the outer shell decreases as the intensity of the main shell decreases 
during the evolution (the analysis for
the different masses is discussed in $\S$ 4.2.3). This brightness ratio
does not decrease steadily  because the formation of
the hot bubble causes a sudden increase in the main shell brightness.
The evolution
of these low brightness outer shells is only affected by the 
ionization of the gas in the time the PN evolves, which is short
compared to the evolution during the AGB phase. The fast wind is not relevant
in their evolution since it only 
affects the innermost regions of the gas structure. The photoionization
generates pressure gradients  
which broaden the outer shells and smooth out the density
structure on large temporal scales. The \ha emission brightness
of the outer shells decreases with radius and shows
a limb brightened edge. Therefore these shells can be interpreted as
detached halos. On the radial scales shown intermediate
detached and attached shells are also present although they are difficult to
see.

In the  5 \Mso~model, the outer regions are not ionized and therefore only
a single shell is formed. A neutral dense shell is also present (see
the density profile in 
Fig. 16) at $\sim$2 {\rm pc}. Although this neutral shell contains a huge
amount of material, we do not expect it to be detectable in H$_2$,
as it is very far away from both the CS and the photodissociation region
for any excitation mechanism in H$_2$ emission to be efficient.

\subsection{The Different Shell Formation Processes}

\subsubsection{The Main Shell}
In a PN, the inner shell is the main observable feature because
of its brightness,  so it is the most widely studied region. The properties of
the main shell change from nebula to nebula, independently of
morphology. Since the energetic input of the CS  directly determines the
evolution of this shell, it is very likely that its radius and expansion
velocity are directly related to the evolutionary status of the nebula and
to the mass of the stellar progenitor.

According to our models the main observable feature of the PN during its
early evolution is that associated with the propagation of
the IF. The total time
that this shell is the main observable feature depends on
the progenitor mass, and is $\sim$3000 {\rm yr} for the models with masses of
1, 1.5, and 2 \Mso, more than 5000 {\rm yr} for the model with 2.5 \Mso,
$\sim$1000 {\rm yr} for the 3.5 \Mso model, and during the whole evolution
for the 5 \Mso~model. We find excellent agreement with the analytical
predictions of \cite{Bk:90}, who found that during the first 2000--3000 {\rm
  yr} of evolution the ionized region is the dominant feature of the PN. 
Afterwards, the evolution of the main shell depends on the 
kinetic energy provided by the hot bubble. The temporal evolution of the
radius and expansion velocity of the main shell 
is discussed in Section~4.1.

\subsubsection{The Intermediate Attached and Detached Shells}
With the exception of the 5 \Mso~model, attached and detached intermediate
shells, that surround the main innermost shell, are formed for all the
models. Detached and attached intermediate 
shells have different formation processes. 
The detached intermediate shells are formed by the effect of the propagation
of the IF. The main shell is that
formed by the IF until the hot bubble develops. When the hot bubble
develops a shell brighter and closer to the CS is formed (becoming the main
shell) and the shell that
was formed by the IF becomes secondary in brightness with a EBP that
decreases more steeply than $r^{-2}$ and show a limb-brightened edge, i.e., a
detached intermediate shell.  

The attached shells, which are characterized by a linear decrease in EBP, are
a consequence of the previous radial density structure. In contrast to what
has been suggested previously \citep{Mel:94,Setal:97}, we find that the
attached shells are not necessarily formed by the dynamical effects of the
photoionization of the gas, since they still appear in the artificially
computed EBP of test models in which the photoionization was switched off. 

The attached intermediate shells show almost  
constant velocities around 20 \kms, and the detached shells,
since they have been accelerated by the IF, have velocities up to 10 \kms\
larger than those of the attached shells. With time, the detached intermediate
shells may not be discerned from the attached shell. 
Guerrero et al. (1998) determined the expansion velocities of a sample of PNe
that present multiple shells. They found that the average expansion velocity 
of the attached shells in their sample was 25 \kms, which is in agreement
with our results. 

\subsubsection{The Detached Halos}
The models computed at larger radial scales show that a faint shell
surrounding the main and the intermediate attached and detached shells
is present. We refer to these shells
as outer detached shells or halos. The imprint of the wind history
experienced by the star is mainly present in 
this outer detached shell. Although the emission
arising from the outer parts is faint compared to the emission from the
main shell, deep optical CCD images should reveal the existence 
of the outer shells, as long as these are ionized. 

The outermost detached shell is a consequence of the previous
density structure, i.e., the shell formed during the AGB phase, and its size
is mainly controlled by stellar evolution, although the ISM pressure also
plays a role (see Paper I, where we showed how the signatures of the
modulations in mass loss consequence of the thermal pulses during the AGB
are not recorded in the gas structure). The shells
formed by shocks in the 
interaction regions between two subsequent periods of enhanced mass loss 
are the only remaining stable 
structures. These are the shells visible during the PN stage in
the form of faint external detached shells or halos (note the huge difference
in brightness between the main and outer detached shell in Figs
11--15). According to our models, the dynamical times derived from
observations of PN halos cannot be correlated with the theoretical timescales
between thermal pulses. A detailed analysis of the
evolution of brightness ratio, velocity and radius of these shells is
presented in Section 4.2. 

\section{DERIVED OBSERVABLE PROPERTIES}
Comparing the results of the models with observations requires some caution. 
Depending on the emission line observed, different linear sizes
and expansion velocities are measured.
In many nebulae there is a difference in the
expansion velocity measured in low (\ha, \nii, and [O~{\sc ii}]) and high
excitation (\heii, \oiii, etc.) emission lines (this is
the so-called Wilson effect [Wilson 1948]), with the low excitation lines
having the largest expansion velocities.
Moreover, most of the 
velocity determinations in PNe have only been obtained for the brightest inner
regions, and all the distance-dependent parameters suffer uncertainties 
because of the poorly determined distances in Galactic PNe.  
Kinematical studies of outer shells are scarce and also have the problem
that their receding and approaching velocity components are unresolved.
\cite{Cja:87} offer a complete discussion on how  distance and  nebular
evolution hinder the detection of faint outer shells in PNe. 

From our models
we have derived properties that can be used  
to compare with the observations. Since we
have assumed spherical symmetry in our models the results 
and the discussion hereafter exclude  comparison with PNe having bipolar
morphology. The detached outer shells of PNe are
mainly detected in \ha or \oiii; no emission is usually found in \nii. All the
analysis from now onwards is made in the \ha recombination line. 

We should point out here that our models assume a homogeneous ISM as well
as an smooth AGB gas ejection. PNe shells are far from being homogeneous 
and very often show different kinds of gas condensations, cometary tails,
and low ionization emission regions \citep{Betal:98}. Different
mechanisms may be responsible for the formation of these small scale
structures in PNe; i.e. instabilities, AGB, and ISM clumps that may have
important consequences on the gas dynamics \citep{Detal:00}. 
\cite{Detal:89} suggested that the clumps are formed in the envelope of
the AGB star and studied 
the conditions needed for its survival to the PN phase. The clumpy nature
of AGB winds has been widely observed in molecular lines \citep{Oetal:00}. A
clumpy AGB wind ejection alters the structure of 
the shocks (cf. \citealt{Wd:02}), and if the clumps are massive
enough to survive the harsh environment they will probably act as broken
walls through which the flow will go (with a modified
density and velocity). It is possible that the main shell formation
will be influenced by the clumpy nature of the AGB wind, but not the outer
shells, as it is very unlikely that a massive condensation can travel so far
away from the star to directly influence the halo formation. We think that
the development of instabilities in the flow is a more plausible explanation
for the presence of small scale structure in the PN halos. We expect 
that only the external shells of the high 
mass progenitors (3.5 and 5 \Mso) may be affected by a a clumpy nature of
the ISM since only in these models is the halo formed
by the compression associated with a shock between the AGB wind and the ISM
material. 
 
It is very likely that these mechanisms are operating all together in
real PNe, and until detailed 2D numerical simulations that take into
account all of this effects are performed we can only speculate on the
relative importance of one or another. 

\subsection{The Main Shell}
\subsubsection{The Evolution of the Radius} 
We have computed the evolution of the main shell radius from the
position of the maximum of the \ha brightness profiles (see Fig.~17). The
position of the inflexion point in the curves of Fig.~17 is 
determined by the point where the shell produced by the impinging of the hot
bubble evolves into
the brightest shell. Before this point, the brightest shell is formed
by the compression associated with the IF. 

\subsubsection{Kinematics}
In order to study the kinematical evolution of the shells in our models, 
we have computed the \ha synthetic line spectrum. The intensity, $I(x,v)$, 
measured
at a given projected distance, $x$, for a velocity $v$ can be computed by
convolving each value of the component of the velocity
vector along the line of sight, v$_{y}$, with the emissivity
function, $\epsilon(r)$, and with the velocity distribution function
$\varphi~(v,T_{\rm e},v_{\rm y})$. The
velocity and emissivity radial distributions are taken from the simulations,
and we have assumed that the electron temperature has a constant
value of 10~000 {\rm K}. 

We have used a spectral resolution of 1 \kms. In the
computed line intensity velocity profiles through the center of the nebula,
the receding and approaching components arising from the higher density
region overwhelm the emission of the other shells. The expansion
velocity of the brightest shell is calculated directly from the
line intensity velocity profiles. 

We have computed synthetic \ha spectra for the photoionized gas at each
output of the simulations, and have derived the 
expansion velocity of the brightest shell. The temporal dependence 
of the main shell 
velocity is shown in Fig.~18 for the first 10~000 {\rm yr}
of the PN evolution. The temporal resolution of
the curves depends on the model and is typically between 135 and
400 {\rm yr}. In all  cases, the data have been smoothed
using a window of width twice that of the resolution element. 

All the models show expansion velocities $\sim$20 \kms\ during the early
stages when the dominant feature is the shell caused by
the IF. The shell is accelerated when the density of the
innermost shell, forced by the thermal pressure provided by the hot bubble,
is larger than the value of the shell formed by the passage of the IF.
From then onwards, this shell is driven at the expense of the
thermal pressure of the hot 
bubble, and its velocity increases rapidly with time. 
As the CS evolves towards lower effective temperature, the wind kinetic energy
decreases (see Fig.~3), and, moreover, the adiabatic expansion of the hot bubble
decreases its thermal pressure. These two processes combined together explain
why the velocity of the main shell is slowed down at later stages during
the evolution.

\subsubsection{Radius versus Expansion Velocity} 
In our models neither the expansion velocity of the nebula nor
its radius increases monotonically with age. 
With the aim of exploring a possible correlation between these two observables
we have plotted in Fig.~19 the radius and expansion velocity of the
inner shell at each time we have an output from the simulations.  We show
the relation for a maximum radius of 0.5 {\rm pc}. The total
time shown is therefore different in each of these panels as it depends on
the time the model needs to reach this radius (see Fig. 17).

Since in Fig. 19 the time is the hidden variable (for each time we plotted
pairs of radius and velocity of the inner shell) 
more than one expansion velocity may correspond to one value of the
radius  (i.e. for the 1 \Mso star).
The radius of the inner shell (defined as the maximum
in the emissivity) for this 
model increases with time at the 
very beginning of the evolution, then decreases (with the rarefraction wave
caused by the IF) and later it is more or less constant until the moment a
hot bubble is formed and takes control of the inner shell (see Fig. 17). 

For the intermediate cases (2, 2.5, and 3.5 \Mso), Fig. 19 mainly
reflects the temporal evolution of the expansion velocity of the
inner shell (see Fig. 18) because of the almost monotonicall increase of
the radius of the inner shell with time for these models (see Fig. 17). 

We find low velocities at small radii for all the models. For
radii larger than $\sim$0.2 {\rm pc} 
only the 1.5 \Mso~model shows a monotonic increase in velocity with
radius. In the 2 and 2.5 \Mso~models, the
velocity increases with radius up to $\sim$0.25 {\rm pc} decreasing
afterwards. In the 5 \Mso~model the velocity remains almost constant around
20 \kms\ as the radius 
increases. The same behavior is shared by the 3.5 \Mso~model but after a
velocity increase with radius up to $\sim$0.1 {\rm pc}. 
The evolution of the energetic input of the CS explains the behavior at
larger radii. The more massive nuclei fade faster in the HR diagram, the hot
bubble is depressurized by its adiabatic expansion, and the main shell
velocity decreases, even though its radius is still increasing. The behavior 
at small radii (i.e., a young nebula) is related to the IF. 

The observed main shell sizes and expansion velocities in PNe are spread
over a wide range, with a tendency of the expansion velocity to
increase with radius up to 0.2 pc \citep{Sab:84b, Bia:92} while for larger
radii the scatter is very large. Our models for different masses show the
same tendencies. As suggested by \cite{Sab:84a} and \cite{Wei:89} in order to
explain the observed scatter of the expansion velocity versus radius in PNe,
we find that the behavior of the
expansion velocity versus the radius depends on the evolution of the CS
energetics and subsequently on the progenitor mass. 

\subsubsection{Kinematical versus CS Ages}
The most common way of deriving the evolutionary status of a PN is through
its kinematical age (the ratio between the radius of the shell and its
expansion velocity). 
Obtaining ages from empirical determination has two fundamental problems.
First, they are subject to the distance uncertainties in determining the
physical size of the nebula. Second, as we have shown previously,
the nebular shells are subject to acceleration during the PN evolution. 

It has been widely assumed that the kinematical ages can be compared with the
evolutionary timescales of the CS.  However, there is some disagreement when
this comparison is made. \cite{Mcetal:90} found a
strong trend for old nebulae to be located at the position of young
CSs. When the sample is biased towards low mass progenitors the
tendency is in the opposite sense (\citealt{Gath:84}, and some low-mass
objects in the sample of \citealt{Mcetal:90}) and young nebulae appear to be
hosted by old CSs. In order to investigate this observed disagreement between
the kinematical and CS timescales, we have computed the kinematical age of the
nebula derived from our models and we plot it in Fig.~20 versus the age of
the CS. 

We find that the kinematical ages are
always higher than the CS ages when the nebula is younger than 5000 {\rm yr},
whilst for intermediate ages (between 5000 and 10~000 {\rm yr}) the ages
derived 
from a dynamical analysis  
tend to overestimate the age of the CS for high mass progenitors (3.5 and 5
\Mso) and 
underestimate it for the low mass progenitors (1, 1.5, 2, and 2.5
\Mso). Thereafter, the 
tendency is maintained except for the models with 2, 2.5, and 3.5 \Mso, for
which the kinematical ages agree well with the CS ages.

This result can be explained by the fact that during the early stages of the
evolution the main shell is a consequence of the IF. 
The radius is larger during early stages of the evolution than later on
when 
the hot bubble shapes the main shell (see Figs~5 and 6). Moreover, the
velocity of the shell 
is slower at the early times when it is determined by the IF, making
the kinematical age even higher.  
Later on, the situation changes; when the hot bubble 
drives the main shell, the radius becomes smaller 
and the expansion velocity increases making the kinematical ages
decrease (1, 1.5, 2, and 2.5 \Mso). The tendency of the kinematical ages
to be higher than the CS ages is maintained for the models with 3.5
and 5 \Mso~because of the fast CS evolution for the former and the fact that
the shell is always driven by the IF for the latter. The evolutionary
timescales of the CS are reflected in the dynamical evolution of the shell. 

With the results of our models we find a natural explanation for the 
disagreement between the kinematical timescales in PNe and the evolutionary
status of their CSs that is able to account for the observed
characteristics. For young CSs the dynamical ages tend to overestimate
the CS ages for all the models. At later stages of the evolution, when the
sample is biased towards low-mass 
progenitors, the dynamical ages underestimate the evolutionary status of the
CSs. \cite{Mcetal:90} found transition times to be the most reasonable
explanation for this disagreement. However, we find that the details of the
evolution of the main shell for the different masses account for the
difference in the observed timescales. 
\subsection{PN Halos}
\subsubsection{The Main-to-Halo Shell Brightness Ratio} 
The main nebula and the detached halo\footnote{The discussion hereafter 
excludes the 5 \Mso\ model since it is ionization-bounded and therefore does
not show a halo.} evolve on different timescales.  
The former is subject to the strong interaction with the fast 
stellar wind, so it expands more rapidly and its H$\alpha$ emissivity 
decreases on a shorter timescale.  With the purpose of exploring the 
evolution of the brightness ratio between the main nebula and the 
detached halo, we have computed the ratio between the peak of the 
surface brightness of each shell and show its temporal evolution 
in Figure~21.  This ratio does not decrease continuously 
during the evolution because the density of the main shell reaches a peak
(when pushed by the hot bubble) causing the maximum seen in the brightness
ratios. 
 
The surface brightness peak ratio between the main nebula and the detached
halo of a sample
of multiple-shell PNe observed by Guerrero et al. (1998) covers a wide
range. There are a few apparently very 
evolved nebulae (IC~1295, M~2-2, NGC~6804, and PM~1-295) in which the
detached halos are only 2--5 times fainter than the bright main nebulae. This
ratio is much larger, more than 100 (MA~3, Vy~2-3, and M~2-40) or even 1000
(NGC~6826, NGC~6884, NGC~6891, and NGC~7662) for other PNe in the sample. 
The sample of \cite{Sp:95} also shows a wide range in
brightness ratios. Only one PN in this sample (NGC 6751) has
a ratio below 10 
 and is apparently an evolved nebula (16~300 {\rm yr}), two 
show intermediate values slightly below 100 (M 1-46 [5,700 {\rm yr}; 
\citealt{Getal:96}], NGC~6629 [$\sim$3,000 {\rm yr}], and Cn 1-5 [no CS age
provided]). NGC 2867 (evolving 
through a helium-burning track), and Tc 1 (apparently young) have values
of between 100 and 500. The detached shells in the sample of Chu et al.
(1987), for
which 
the measurements of the main to outer surface brightness peak ratios are
available, are NGC 2438 ($\ge$ 143), NGC 6543 ($\ge$ 667) and NGC 6720 ($\ge$
200), no evolutionary status of the CS is provided.

In our models the extremely low (less than
10) emission brightness ratios are only reached for very evolved CSs. 
Larger ($\ge$ 500) ratios are found only  during the early stages
of the evolution, i.e. for young CSs.
It is clear that the observation of these outer external shells in PNe may be
difficult because of  the huge brightness contrast between the main shell and
the halo. 

\subsubsection{Radius of the Halo versus its Expansion Velocity}
We show in Fig.~22 the expansion velocity of the
halo versus its radius for the different models. The 
points represent each time we have an output from
the simulations. The velocities are computed from 
synthetic line spectra across the
center of the nebula in the same way as explained in Section 4.12. In order to
obtain the expansion velocities of 
the halos we have performed the numerical integration starting from the
position of the radius of the main shell, otherwise the emission from
the main shell completely dominates  the spectra.  

We obtain expansion velocities for the halos that range
between 8 and 17 \kms\ and shell radii between $\sim$1.3 
{\rm pc} and up to 2.2 {\rm pc}. In our models we find that
the outer shells are confined between two shocks caused by the
ionization of the structures. We find that for the 1 and 1.5 \Mso~models
the gas velocity is not constant throughout the shell.

There are very few halos for which the velocities have been measured.
The observed expansion velocities of the halos in the sample of 
Guerrero et al. (1998)
range between 12 and 28\kms, the average expansion velocity being
21\kms. Other individual determinations of the expansion velocity of the
detached halos have been obtained for NGC~2022 (20 \kms) and NGC~2438 (8 \kms)
by \cite{Cj:89}, 
NGC~6720 (25 \kms) by \cite{Gmc:97}, 
NGC~6751 (10 \kms) by \cite{Cetal:91},      
NGC~6826 (13 \kms) by \cite{Betal:92a},   
NGC~6543 (7 \kms) by \cite{Betal:92b},   
NGC~3587 (10 \kms) by \cite{Metal:92}, and 
for M~1-46 (8 \kms) by \cite{Getal:96}.  

Our 
models predictions cover the same velocity range. Although we do not reach
the high expansion velocities  
reported for some of the halos in Guerrero et al. (1998), we have demonstrated
in Paper I how the expansion velocity of the outer shells increases by a
factor of 1.5 on average, when the star evolves in an ISM ten times less
dense.  

\subsubsection{Main-to-Halo Shell Radii Ratio}
Figure 23 shows the  $R_{\rm e}$/$R_{\rm i}$ ratio, where $R_{\rm i}$ is the
main shell radius and $R_{\rm e}$ is the radius of the detached halo versus
the 
radius of the main shell during 15~000
{\rm yr} from the ionization of the halo.
The radius of the outer shell changes
very slowly when compared with that of the main
shell. Therefore the main-to-outer shell radius ratio can be used as an
indication of the evolutionary status of the nebula

All the models fall
in the lower mid-plane of the plot in Fig.~23, following curves that are
inversely proportional to the radius of the main shell since the evolution
of the radius of the outer shell is very slow. Therefore the
curves show an evolutionary sequence, i.e., a 
close correlation with age, that is mainly controlled by the evolution of the
main shell. The curves are displaced depending on the radius of the halo;
the curve for the 1 \Mso~model is the closest to the $y$-axis.
The curves obtained for the models with masses 1.5, 2, 2.5, and 3.5
\Mso~are very close together and are almost indistinguishable.

If we assume that
young PNe have CS ages less than 3000 {\rm yr}, we obtain an main shell
radius smaller than  
0.15 {\rm pc} (see Fig. 17) for which $R_{\rm e}$/$R_{\rm i}$ is greater
than 10 for the 1 \Mso~model and greater than 15 for the rest of the
models. This results in PNe with detached halos more than 10--15 
times larger than their main shells. Obviously, $R_{\rm e}$/$R_{\rm i}$
decreases as the age of the nebula increases (owing to the rapid increase in
the main shell radius). However, it is important to note here that the
detached halos are never less than twice the size of main shells. 
Since the  $R_{\rm e}$/$R_{\rm i}$ ratio is distance-independent, 
the largest source of uncertainty in the observed data comes
from $R_{\rm i}$. As defined by \cite{Kal:74}, giant halo PNe should 
have $R_{\rm e}$/$R_{\rm i}$ greater than 5. According to our models these
ratios are reached only for evolved nebulae, being even larger in earlier
stages.

The outer detached radius in our models is mainly determined by the AGB
evolution of the star. A smaller radius for the PN halo can  be obtained only
if the evolution during the thermal pulsing AGB is shortened. However, it is
important not to forget the possible observational bias towards the detection
of giant halos. As it already pointed out by Chu et al. (1987), the
interstellar reddening at low Galactic latitudes and the intrinsic faintness
of  PN detached halos make them very difficult to detect. 

\subsection{Ionized Masses}

We have computed the ionized mass in the grid and
plotted it versus time in Figure 24. The values for the ionized masses 
while the nebula is optically thick are too small to be appreciable in
this linear plot. When the nebula becomes optically thin to the Lyman
radiation from the star, the ionized mass suddenly increases, reaches a
maximum value, and remains constant from there onwards. 

Previous numerical models have calculated the ionized masses in PNe
\citep{Sk:87, Ms:91, Mel:94}. In Mellema's models the
ionized mass was found to decrease after the nebula becomes optically
thin. This was caused by the small grid size used, which allowed the matter
to leave the grid at the outer boundary. The ionized masses he derived were
as large as 0.22 \Mso. \cite{Sk:87} and \cite{Ms:91} found that the ionized
masses increase with time or radius; however, these results are invalidated by
numerical artifacts in the code they used. 

We find that, after the nebula becomes optically thin to Lyman 
radiation, the ionized mass remains constant with
values that are different for each stellar mass considered.  
Our results do not suffer from the problems that previous numerical
models had in deriving the ionized masses: we have used large enough grids so
the matter does not escape at the outer boundary, 
and in our Eulerian code, the position of the inner and outer boundaries does
not change with time. Moreover, we have followed the evolution of the stars
from the time when substantial mass loss rates take place (the AGB). 

In Fig.~25 (solid line) we have plotted the ionized mass at 9000 {\rm yr}
versus the initial mass of the star. The dashed line represents the amount 
of ionized mass contained within the main shell. It has been
computed by integrating all the ionized mass up to the first emissivity
minimum after the position of the brightest shell. The dashed-dotted line
represent the total amount of mass lost by the star. It can be seen that the
amount 
of ionized mass increases with the mass of the star when the nebula is
density-bounded (stellar cases 1, 1.5, 2, 2.5, and 3.5 \Mso). In the case of
the 5 \Mso~star the nebula remains optically thick to the hydrogen-ionizing
radiation and therefore does not follow the relation of increasing ionized
mass with progenitor mass. 
 
One might have expected  the constant value reached by ionized
mass (see Fig. 24) to be the difference between the initial and final mass of
the progenitor; however, the ionized mass is clearly higher than this value
(see Fig 25). The extra mass comes from the ISM. We showed in Paper I 
that the stellar wind sweeps up a high fraction of ISM matter during the AGB
phase.
This ISM matter swept up by the wind is not at all negligible and resides
mainly in the outer detached shell. The ionized mass
in the detached halos is mainly determined by the mass present in these
shells at the end of the AGB phase since it 
does not increase substantially during the short time the PN phase
lasts. Moreover, the stellar wind during the PN phase does not contribute
significantly to the ionized mass because it has an extremely low
density.

The estimation of ionized masses in PNe can only be made
for objects for which the distances are 
known. The ionized masses observed cover a wide range:
\cite{Bs:94}, using a Galactic sample of 31 PNe, found 
$0 < M_{\rm i} <0.2$~\Mso. \cite{Pot:96} determined the masses for a sample
of 46 nearby PNe, and, with the exception of some objects that have high
masses (4.2, 2.7, 3.5, 2, and 1.8 \Mso~for A~74, IW~2, S~216, WDHS1, and A~7
respectively) most of the objects studied have masses between 0.01 and 0.8
\Mso. Ionized masses have also been determined for some individual objects:
 0.072 \Mso~for M~1-46
\citep{Getal:96}, 1.17 \Mso~for the halo of  NGC 6543 \citep{Mp:89}, and 0.94 
\Mso~for the halo of NGC 6826 \citep{Mcw:89}.

Most of the observed ionized masses lie between 0.1 and 0.25
\Mso. We find that the observed ionized masses agree with the values
obtained in our simulations when only the brightest part of the nebula is
considered (dotted line in Fig.~25). 

It has been claimed that the wide range of observed ionized masses is due to
the ionization-bounded nature of the nebula. In the light of our results we
have to disagree with this. With the exception of the 5 \Mso~star, all the
models result in density-bounded PNe. We think that a more plausible
explanation is that most of the ionized masses were derived from 
the brightest part of the nebula. 

It has also been found that larger masses are derived for lower electron
densities, and that the ionized masses are correlated with the nebular radius
\citep{Pot:84,Bs:94}. These correlations can be understood as an
effect of the evolution. If, as we propose, the observed masses are  taking
into account only the mass contained in the brightest shell, as this shell
evolves, its radius increases, hence 
encompassing more mass whilst its electron density decreases because of
geometrical dilution effects. 

We have to conclude that, according to our models, the observations severely
underestimate the ionized mass that is 
present in PNe since they can  recover only  0.2 \Mso\ on average. Based on a
very simple momentum-conserving two-wind  
model, \cite{Bs:95} demonstrated how standard techniques
severely underestimate the ionized masses in PNe. Selection
effects probably hinder the observation of the 
large low surface brightness part of the nebula, the PN halos, where 
the largest fraction of ionized matter is present.

In Table~2 we have summarized the effective temperatures of the
star for which the nebula becomes density-bounded. These values are
approximate because they are subject to 
time errors of the order of $\sim$250 {\rm yr}, which 
is the average temporal resolution of the outputs in our simulations.
By analyzing the observed effective temperature of 
PN CSs  determined from different methods, \cite{Kj:91}
find that PNe become 
optically thin for temperatures in the range 40~000--50~000 {\rm K}. According
to this, our models provide a very early start for the optically thin phase
for the 1 \Mso\ CSs. 
The radii at which the transition from optically thick to optically thin
nebulae takes place are also included in Table~2. The average
radius is 0.11 {\rm pc}, which agrees very 
well with the radius of
0.12$~\pm~$0.02 {\rm pc} obtained by \cite{Dau:82} for the transition.

\begin{deluxetable}{ccc}
\tablenum{2}
\tablewidth{15pc}
\tablecaption{Optically thick--thin transition radii and CS temperatures}
\tablehead{\multicolumn{1}{c}{Mass [\Mso]}& 
\multicolumn{1}{c}{T$_{eff}$ [{\rm K}]} & 
\multicolumn{1}{c}{R$_i$}[{\rm pc}]}
\startdata
1   & 20,000  & 0.0912\\        
1.5 & 39,000  & 0.0980\\
2   & 130,000 & 0.1029\\
2.5 &128,000  & 0.1783\\
3.5 &207,000  & 0.0681 \\
5   &$\sim$& $\sim$ \\
\enddata
\label{teff}
\end{deluxetable}

According to our simulations, as the
stellar mass increases, higher effective temperatures are needed for the
transition 
from optically thick to optically thin PNe. Moreover, the nebula remains
ionization-bounded longer for higher 
mass  progenitors.

\section{CONCLUSIONS}
We have investigated PN formation considering the gas
structure resulting from the preceding AGB
as the starting point for our models, and using hydrogen-burning 
post-AGB tracks taken from VW94 with core masses 0.569, 0.597, 0.633,
0.677, 0.754, and 0.9 \Mso, and solar metallicity.

As the result of our models we highlight the importance of the dynamical
effects of ionization in 
the shell's evolution. During the early stages of the PN evolution the 
main shell is formed by the IF and is not driven by the hot
bubble. We find that, with the exception of the 5 \Mso~star model, 
multiple ionized shells are present in all the stellar models.
The emission line profiles that characterize the attached 
shells result from a previous density distribution present in the
circumstellar gas prior to the onset of the PN phase. The halos have sizes up
to 2.3 {\rm pc} (more than twice the size of the main shell) and are 
formed during the AGB phase,  with \ha emissivities between 10 and 5000 times
fainter than the main shell. We find that intermediate detached shells 
are formed in our models by the IF.

We have studied the dynamical evolution of the main shell 
and found that for young CSs, the kinematical ages tend to
overestimate the CS ages for all the masses. At later stages of evolution,
the kinematical ages underestimate the evolutionary status of the CSs when the
sample is biased towards low mass progenitors. The details
of the evolution of the main shell for the different masses accounts for the
difference in timescales between low- and intermediate-mass progenitors.

According to our models the observations severely underestimate the ionized
mass present in PNe as most of the ionized mass in PNe is contained in the
the detached halos, which are not usually detected because of their faintness.

%
We thank M. L. Norman and the Laboratory for Computational Astrophysics for
the use of ZEUS-3D. We also want to thank Letizia Stanghellini, Martin
Guerrero, and Tariq Shahbaz for their careful reading of the manuscript and
their valuable comments. The work of EV and AM is
supported by Spanish grant PB97-1435-C02-01. GGS 
is partially supported by
grants from DGAPA-UNAM (IN130698, IN117799, and IN114199) and CONACyT
(32214-E).


\begin{figure}
\plotone{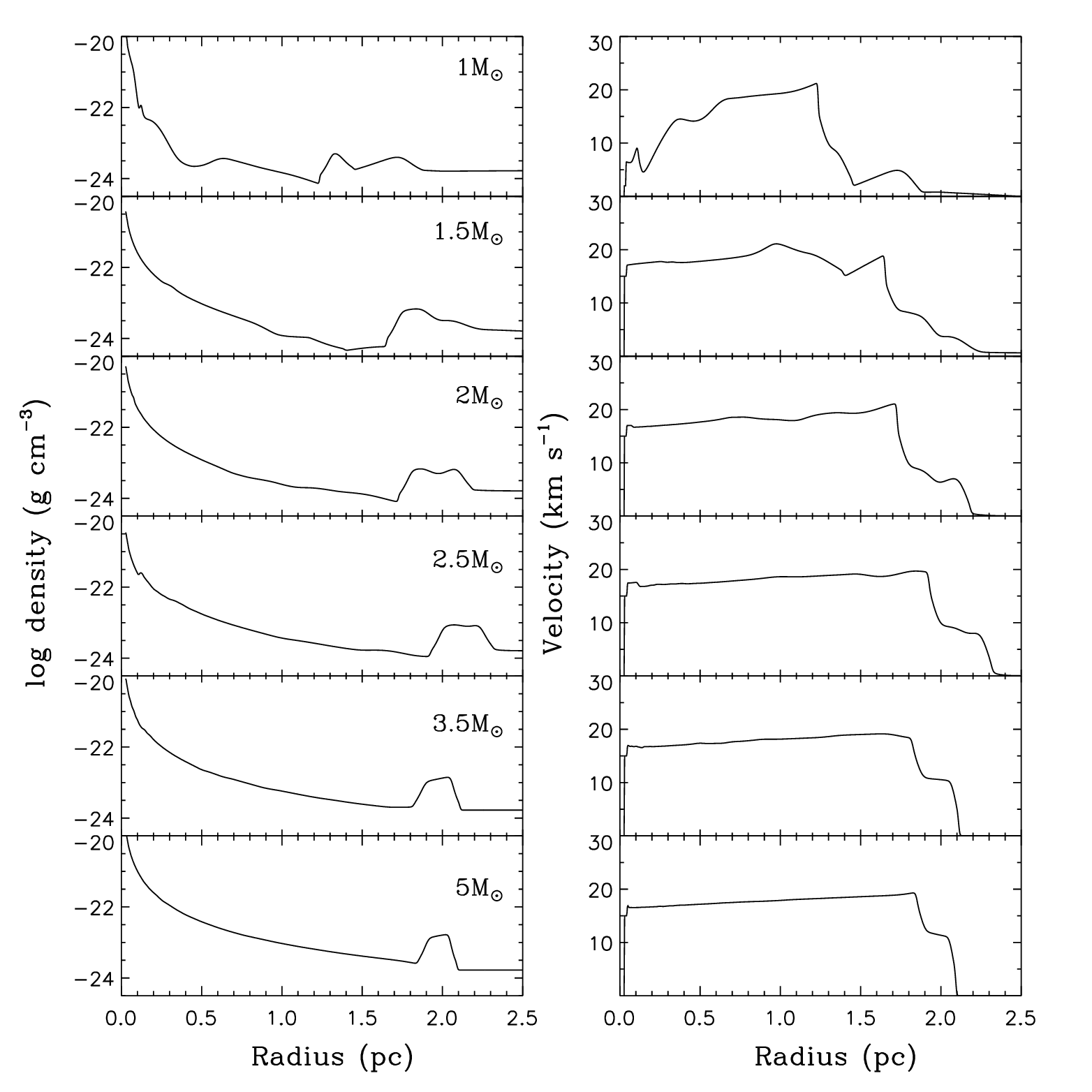} 
\caption[ ]{
     Left: logarithm of the gas density in g~\cm3 
     at the end of the AGB. Right: radial distribution of the expansion 
     velocity in \kms.
     From top to bottom, the initial stellar masses are 1, 1.5, 2,
     2.5, 3.5 and 5 $M_{\odot}$.\label{f1.eps}}
\end{figure}

\begin{figure}
\plotone{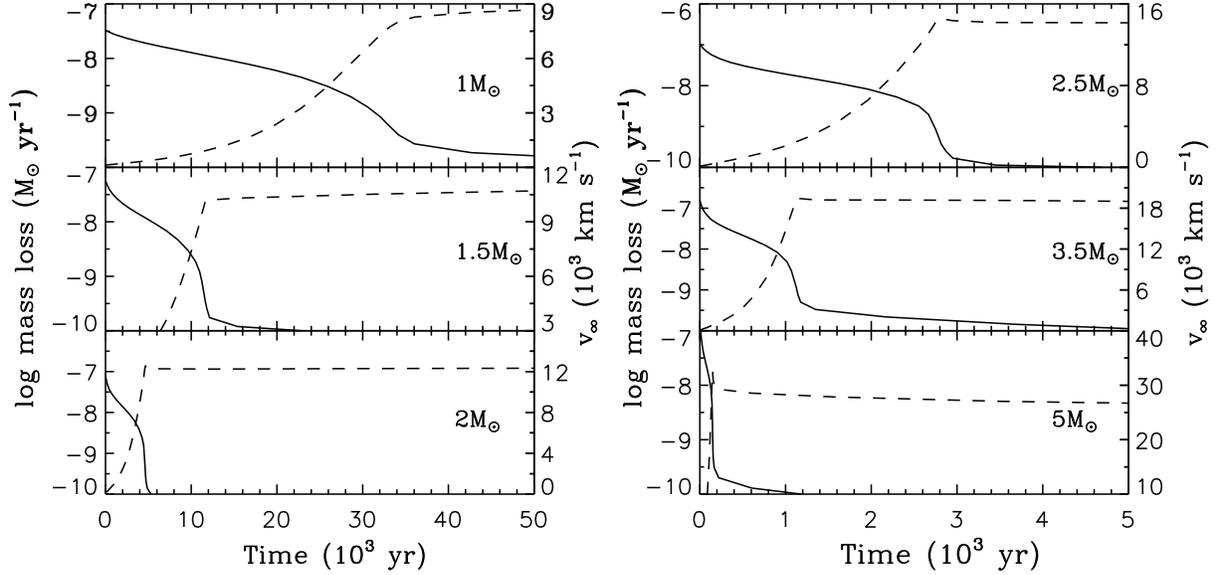}
\caption[ ]{Wind parameters used for our models. 
            Left panels: initial stellar
     masses 1, 1.5, and 2 \Mso. Right panels: initial stellar masses 2.5, 3.5,
     and 5 \Mso. The solid line represents the logarithm of the 
     mass loss rate (left scale) and the
     dashed line the wind terminal velocity (right scale) in units of
     $10^{3}$ \kms. 
\label{f2.eps}}
\end{figure}

\begin{figure}
\plotone{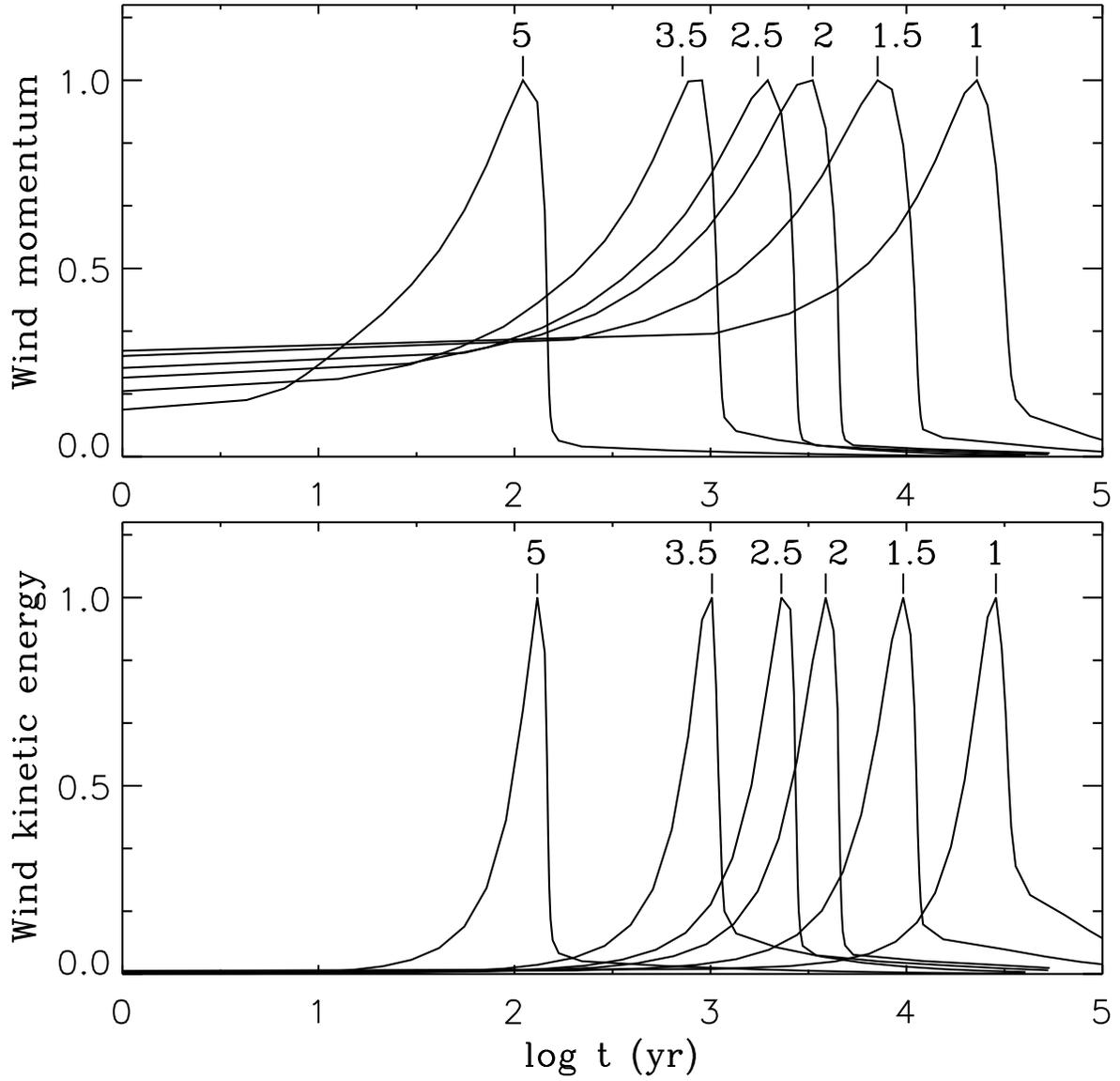}
\figcaption[ ]{Upper panel: wind momentum 
      versus time on a logarithmic scale. Each
      curve has been normalized to its maximum value and marked with the
      initial mass of the star. Lower panel: the same for the wind kinetic
      energy. 
\label{f3.eps}} 
\end{figure}

\begin{figure}
\plotone{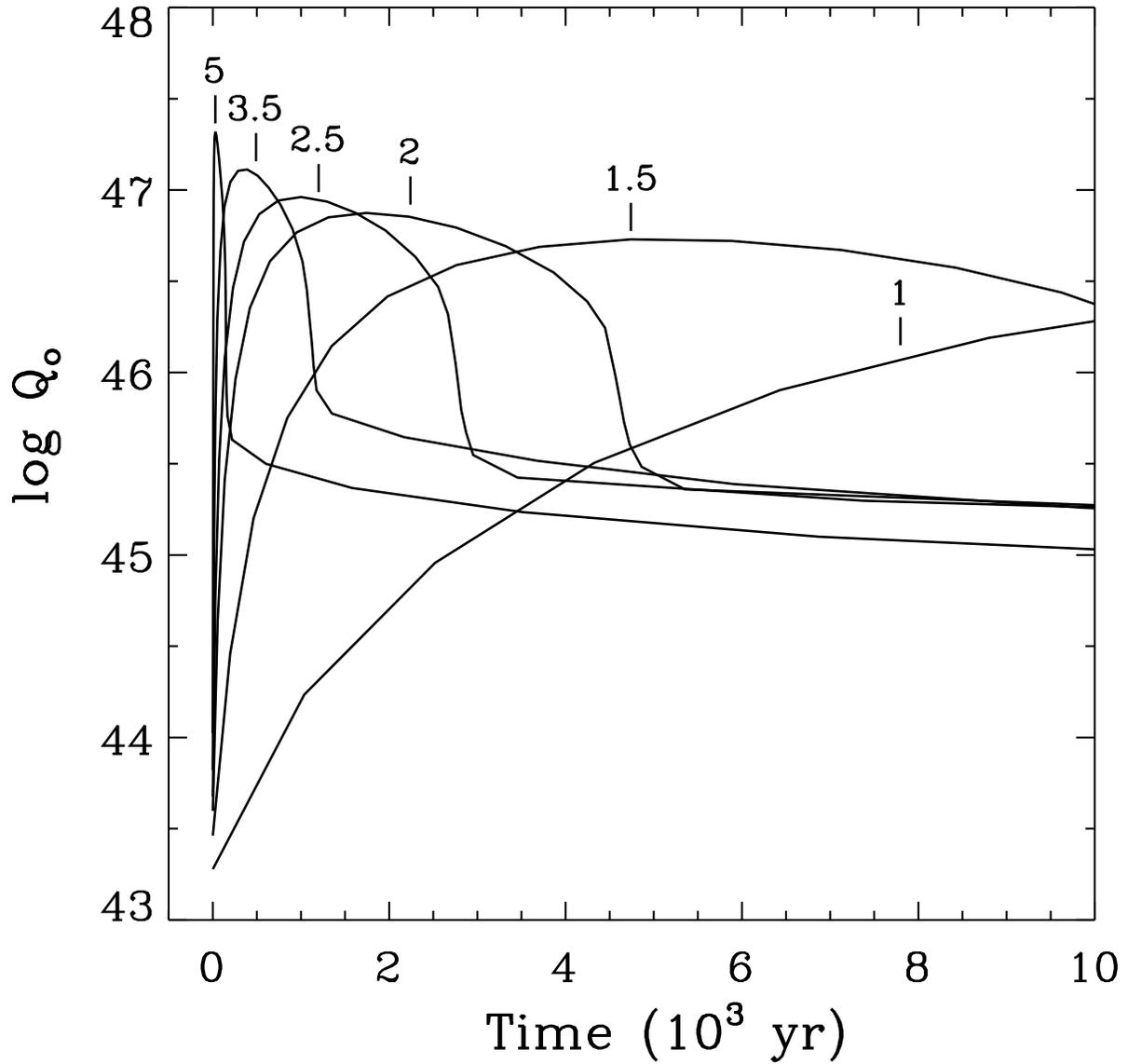}
\caption[ ]{Number of ionizing photons, $Q_o$ (on a logarithmic scale), during
     the first 10~000 yr of  CS evolution. Each line has been marked with
     the initial mass of the model.
\label{f4.eps}}
\end{figure}

\begin{figure}
\plotone{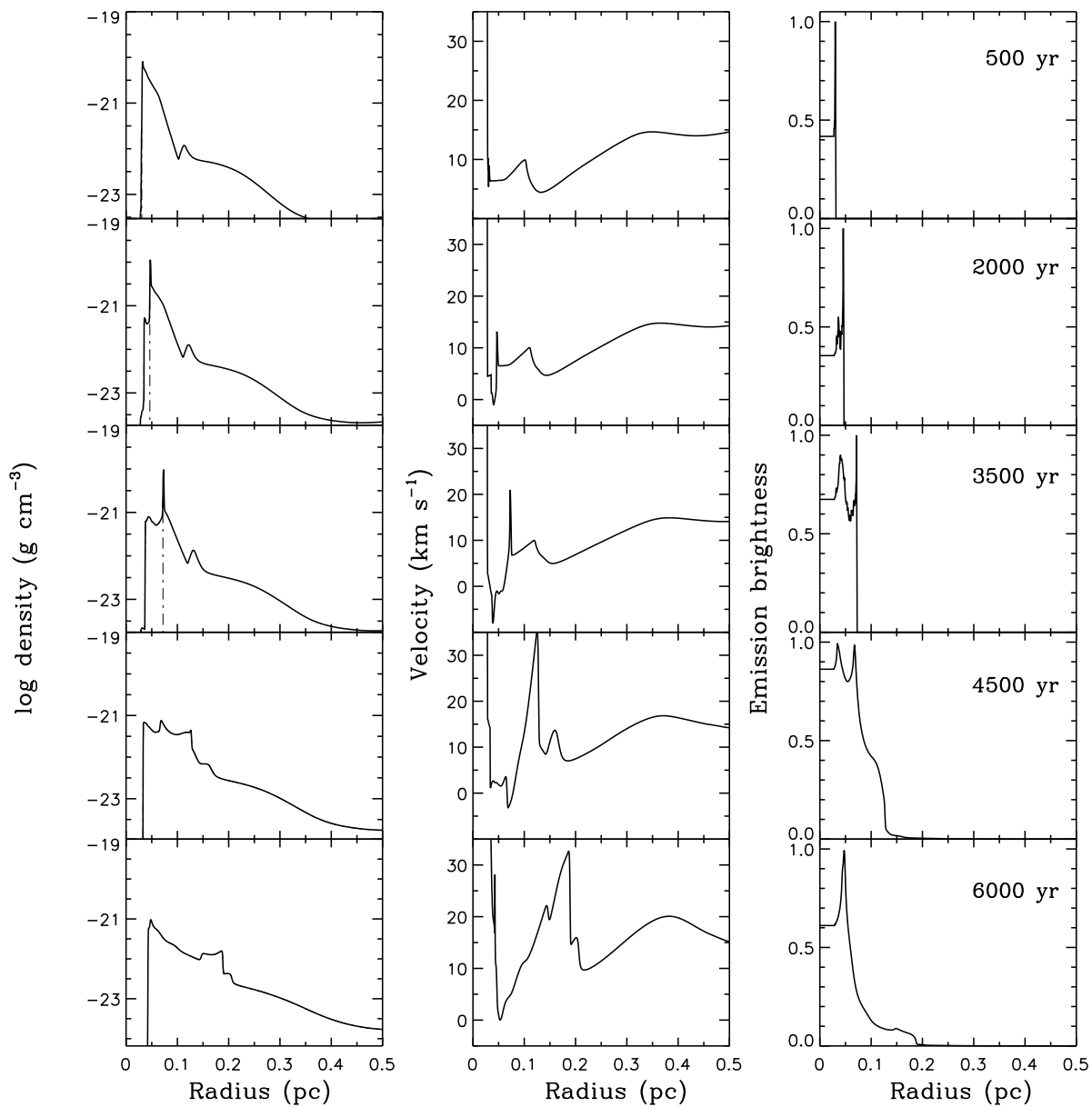}
\caption[ ]{Density, velocity, and \ha EBP
      radial distributions at different evolutionary times during the
      PN stage for the 1 \Mso~stellar model. The position of the IF is marked
       as a dotted line in the density profile.
\label{f5.eps}}
\end{figure}

\begin{figure}
\plotone{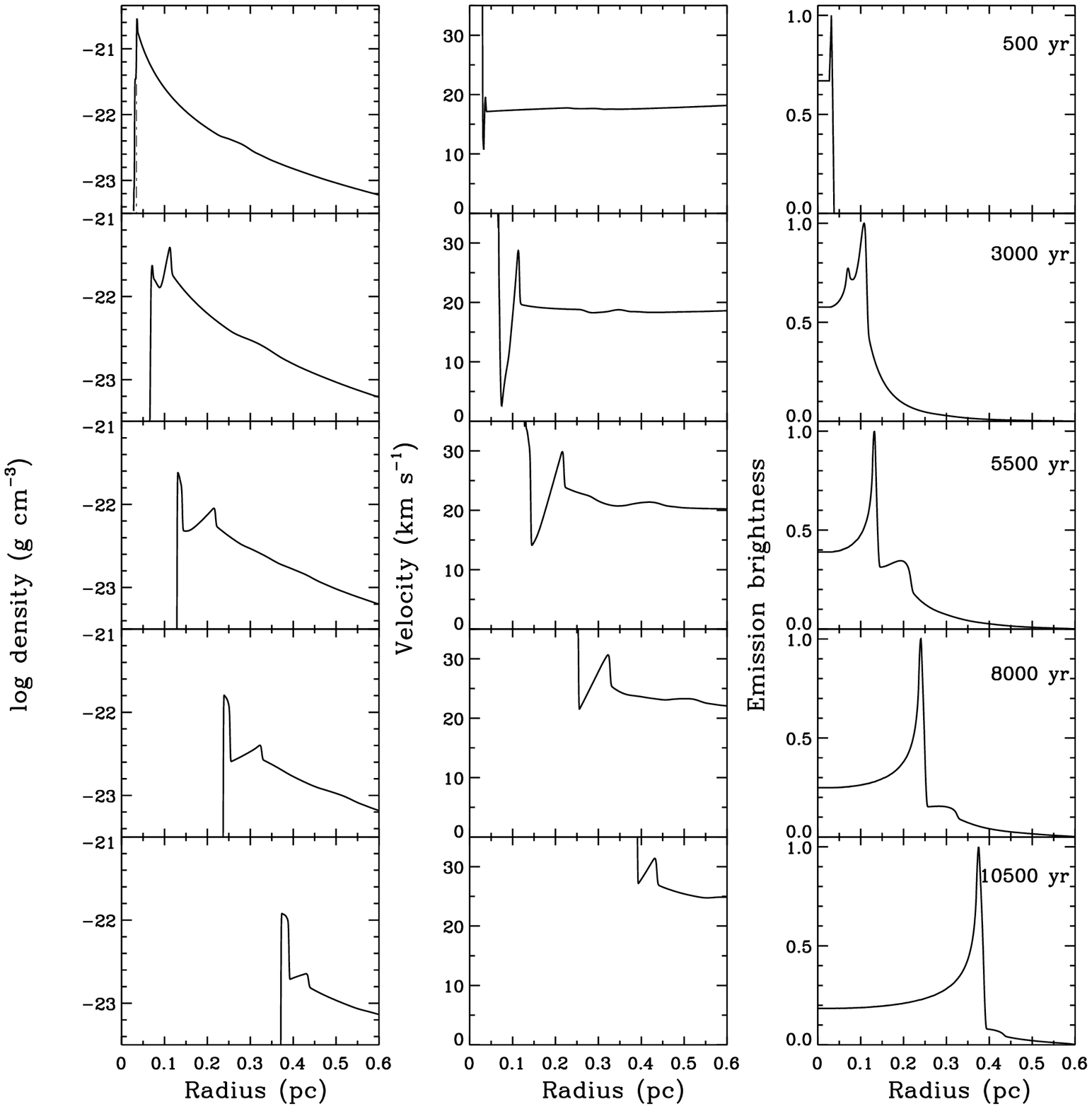}
\caption[ ]{Same as Figure 5, but for the 1.5 \Mso~stellar
      case. 
\label{f6.eps}}
\end{figure}

\begin{figure}
\plotone{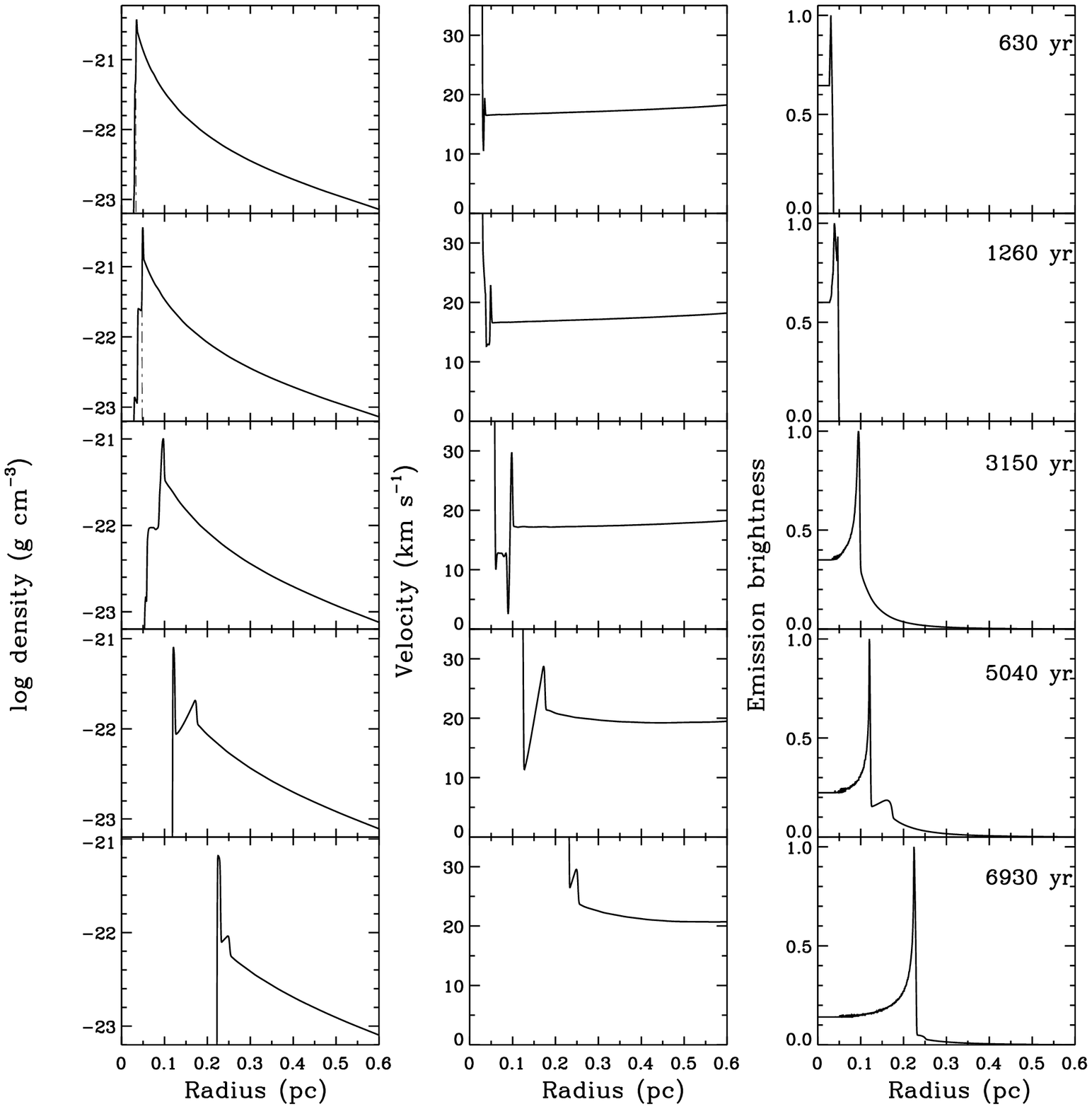}
\caption[ ]{Same as Figure 5, but for the 2 \Mso~stellar case.
\label{f7.eps}}
\end{figure}

\begin{figure}
\plotone{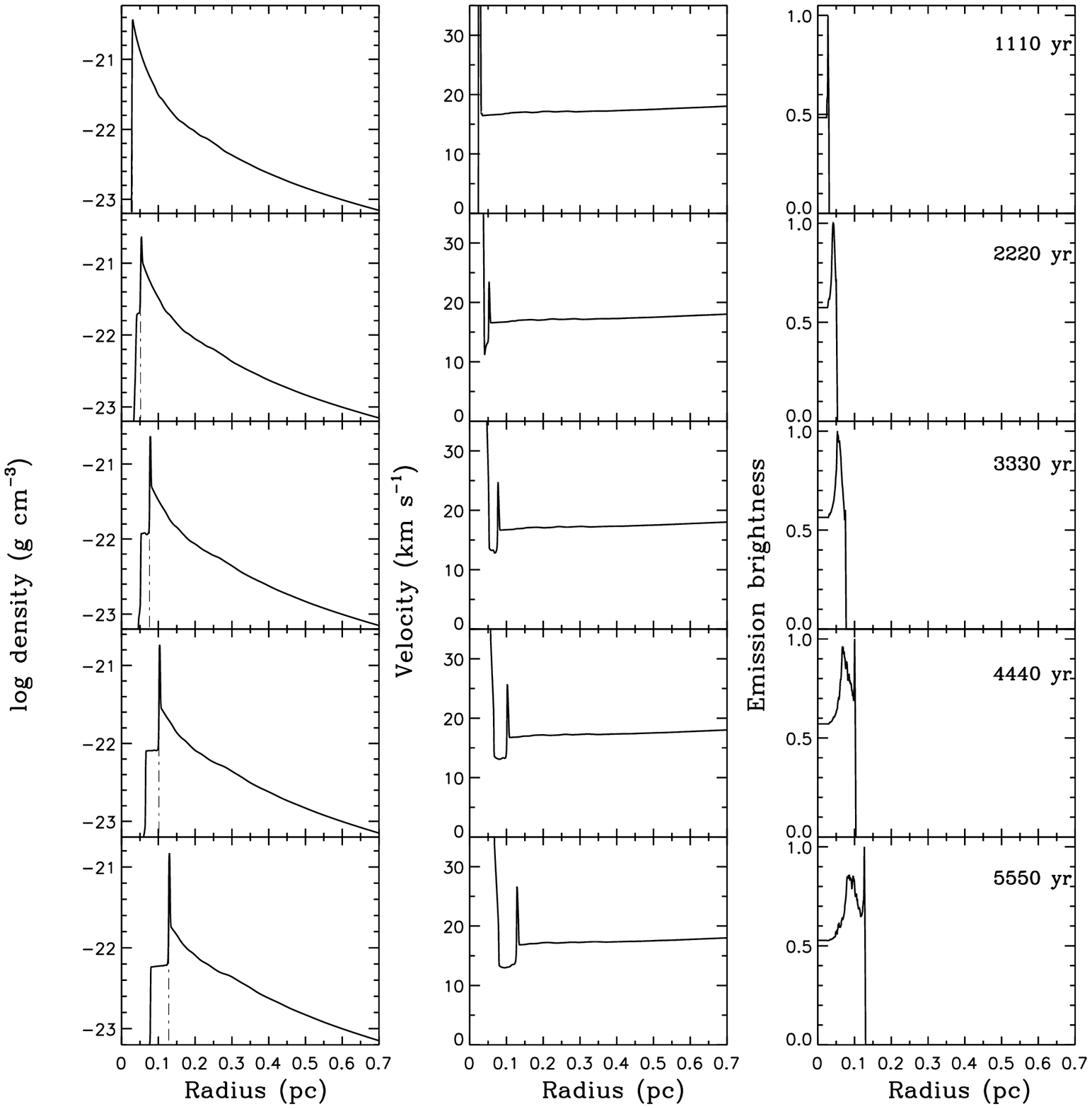}
\figcaption[ ]{Same as Figure 5, but for the 2.5 \Mso~stellar case.
\label{f8.eps}}
\end{figure}

\begin{figure}
\plotone{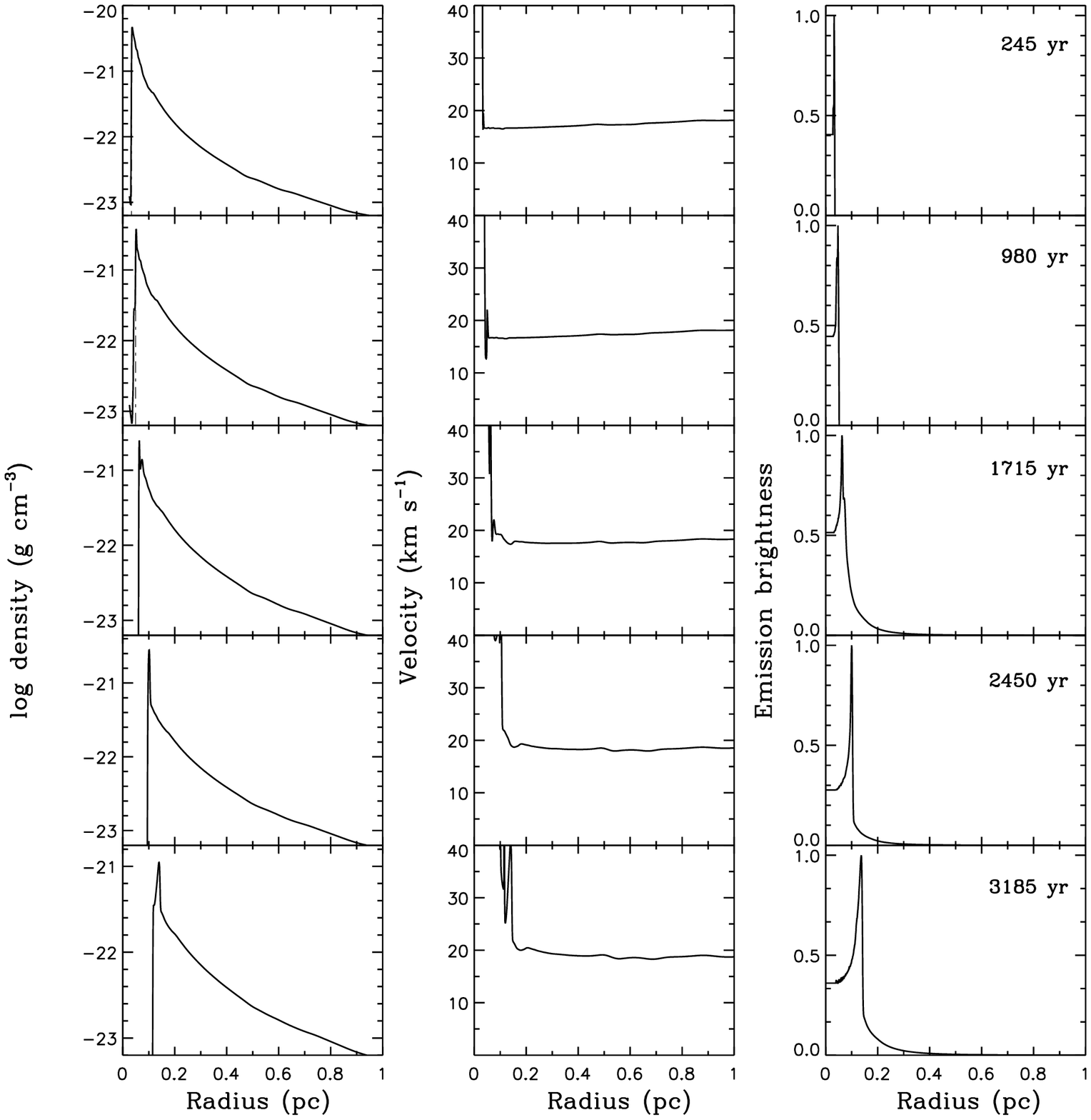}
\caption[ ]{Same as Figure 5, but for the 3.5 \Mso~stellar case.
\label{f9.eps}}
\end{figure}

\begin{figure}
\plotone{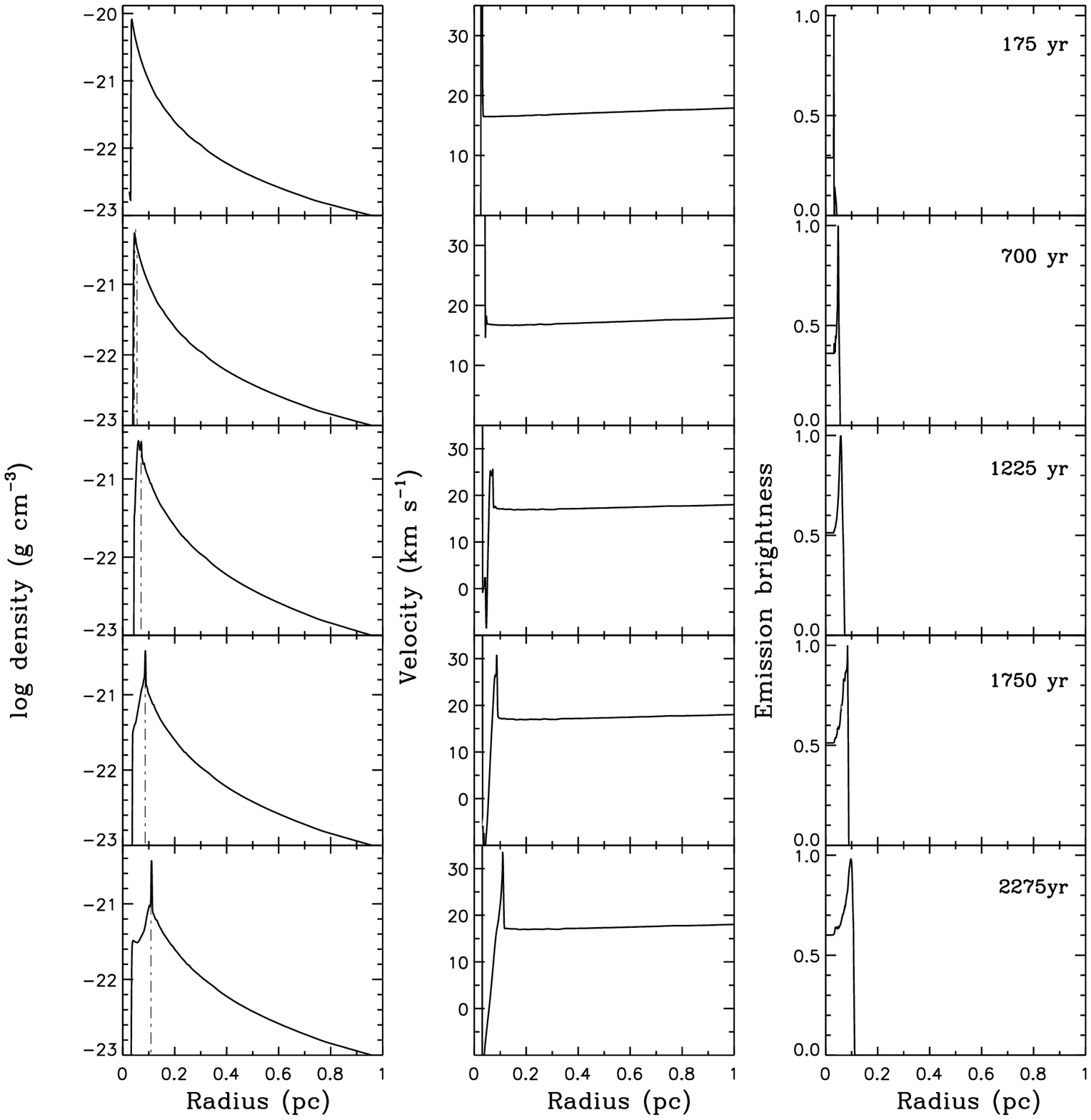}
\caption[ ]{Same as Figure 5, but for the 5 \Mso~stellar case.
\label{f10.eps}}
\end{figure}

\begin{figure}
\plotone{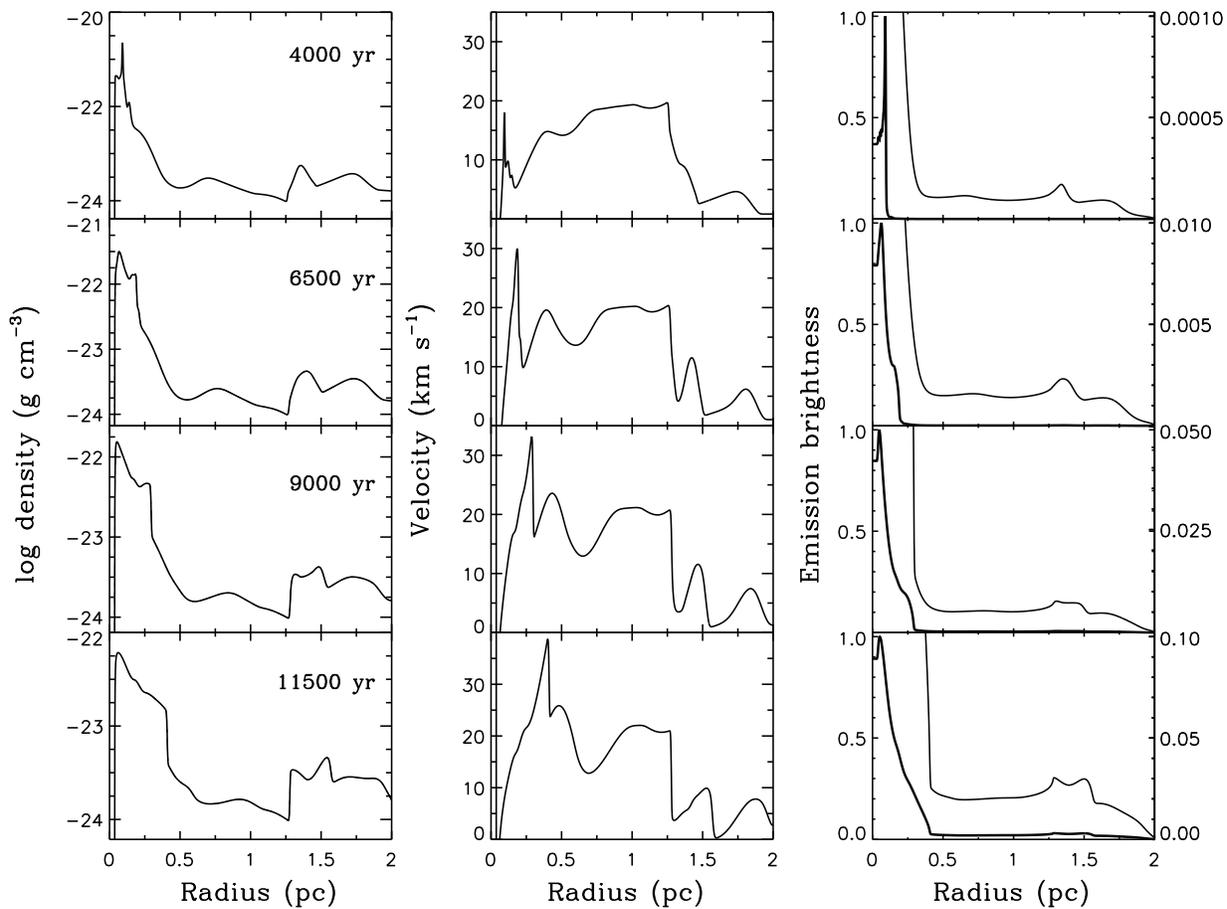}
\caption[ ]{Density, velocity, and EBP radial distributions of the 
large-scale structure of the gas for the 1
  \Mso~stellar case. The EBP are shown on two scales: the scale on the
  left is for the EBP of the bright main nebula, which is shown with a thick
  line; the scale on the right  corresponds to the thin line, which shows
  the EBP of the much fainter outer regions of the nebula. 
\label{f11.eps}}
\end{figure}

\begin{figure}
\plotone{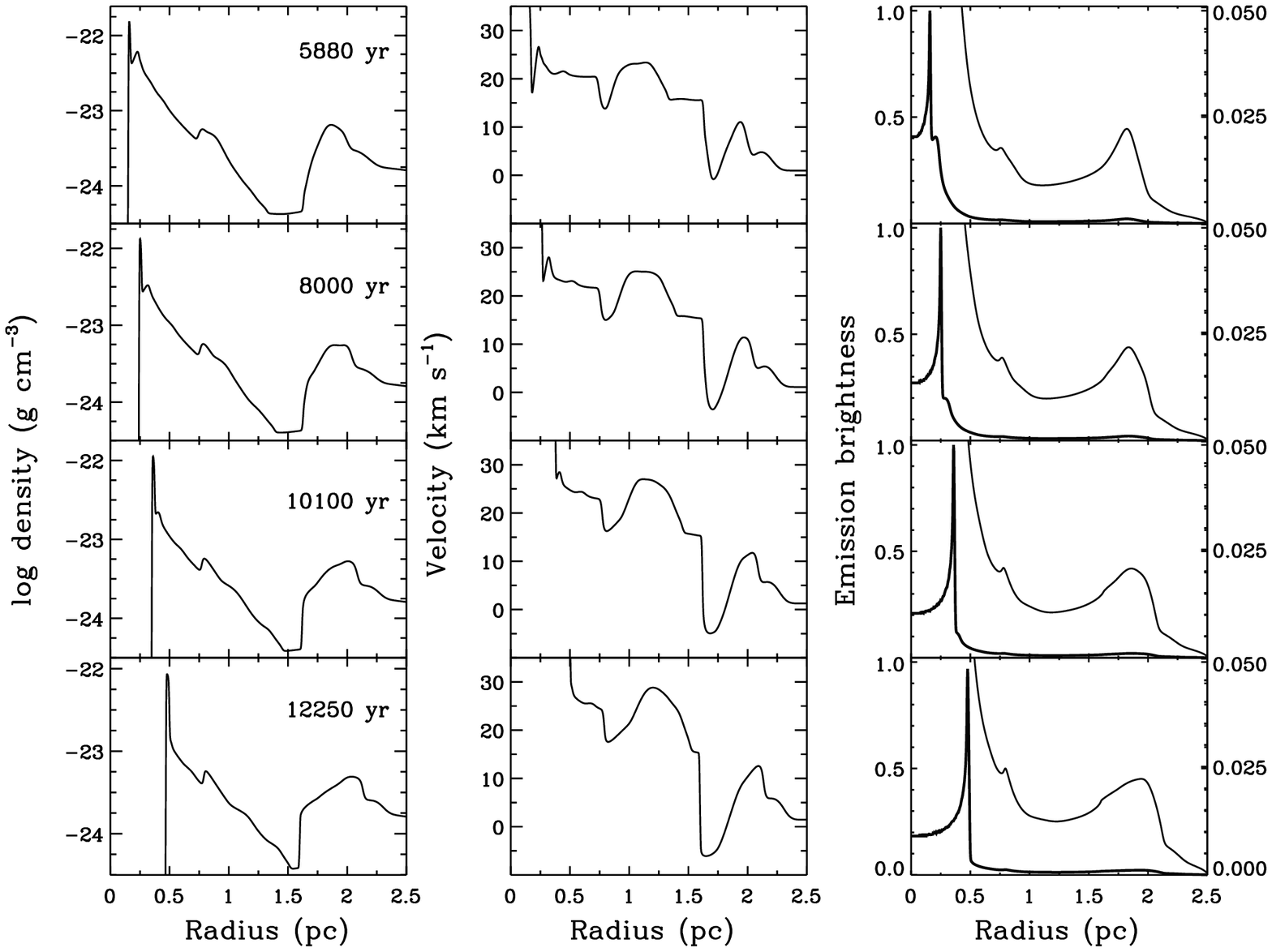}
\caption[ ]{Same as Figure 11, but for the 1.5 \Mso~stellar case.
\label{f12.eps}}
\end{figure}

\begin{figure}
\plotone{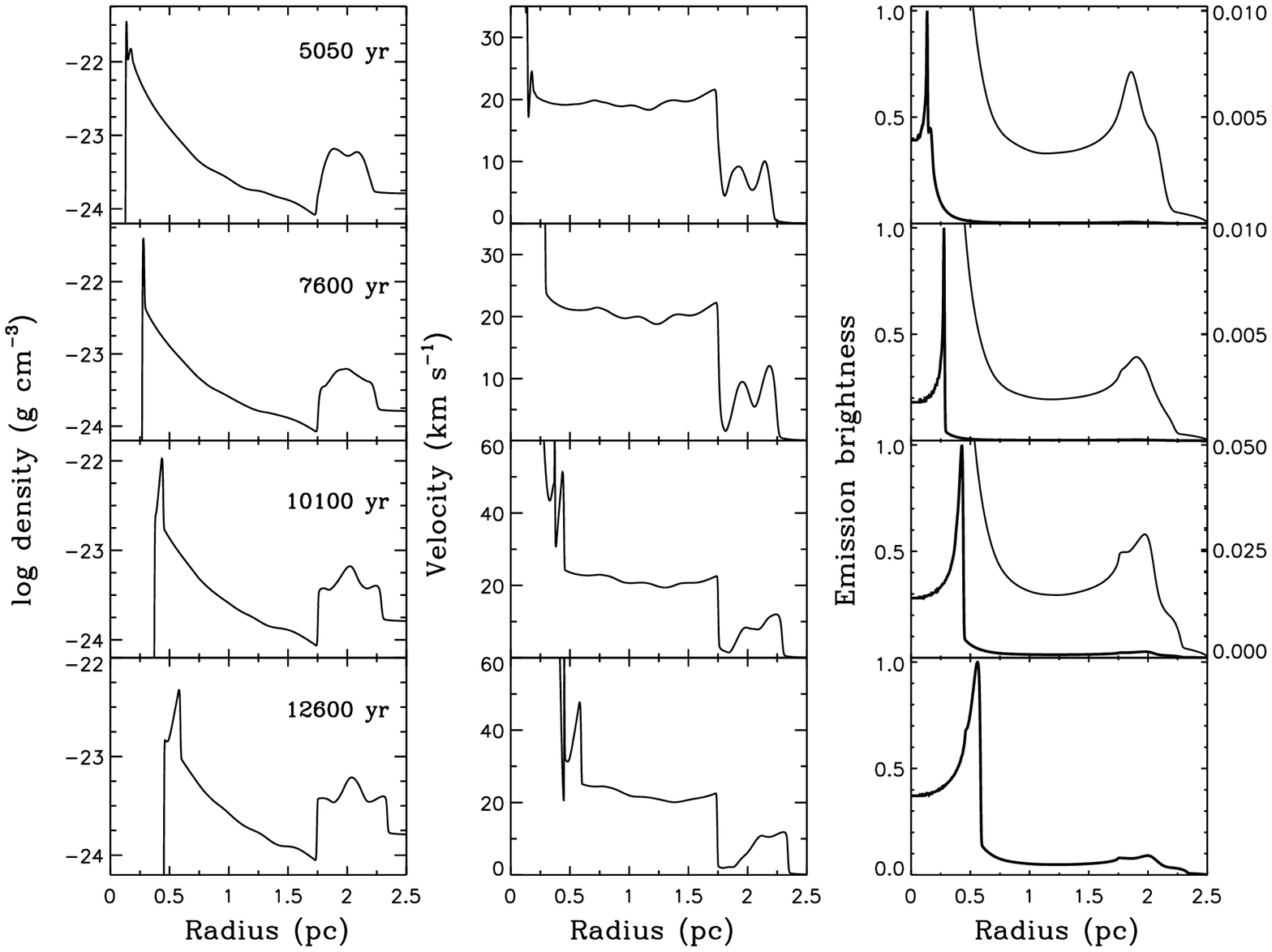}
\caption[ ]{Same as Figure 11, but for the 2 \Mso~stellar model.
\label{f13.eps}}
\end{figure}

\begin{figure}
\plotone{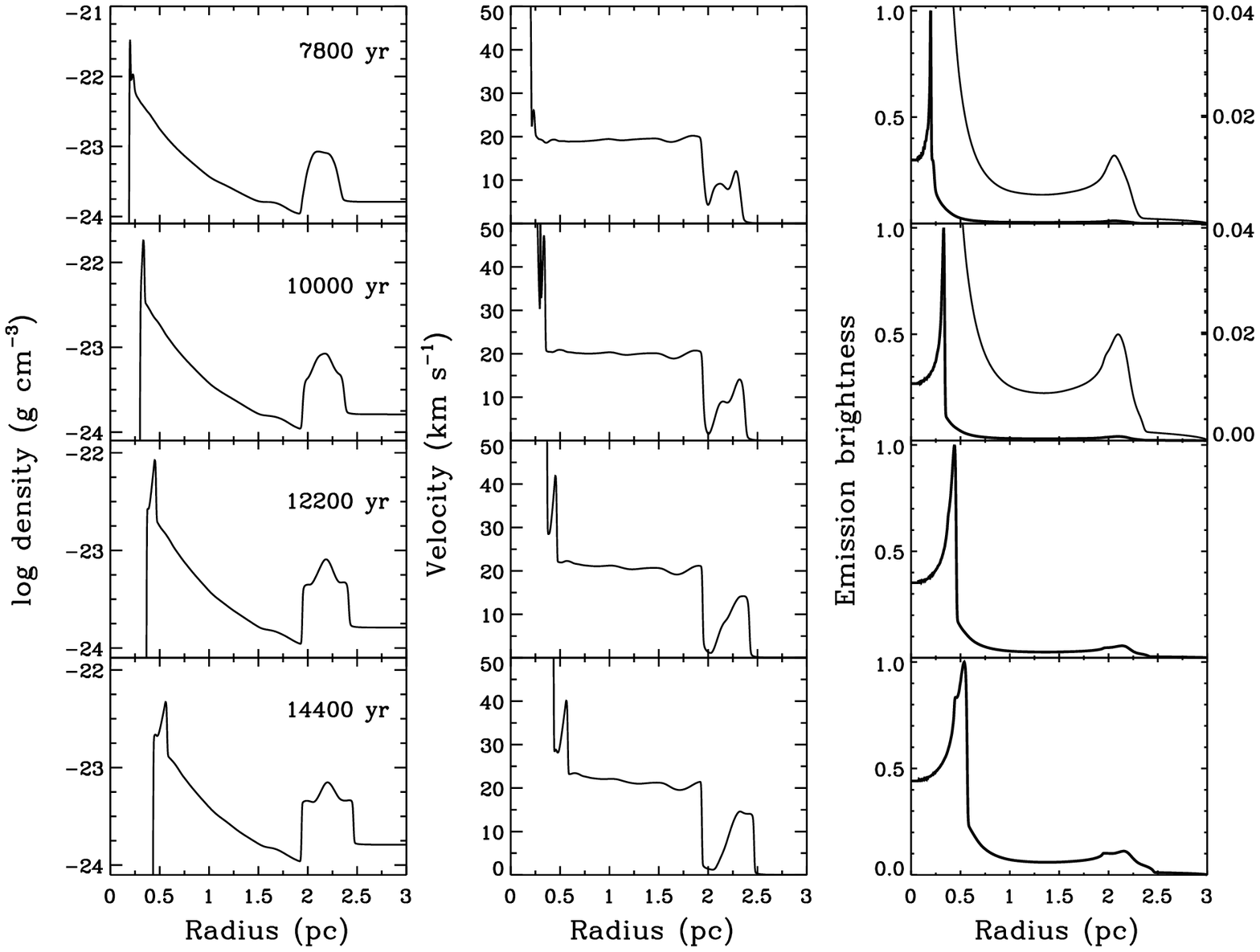}
\caption[ ]{Same as Figure 11, but for the 2.5 \Mso~stellar model.
\label{f14.eps}}
\end{figure}

\begin{figure}
\plotone{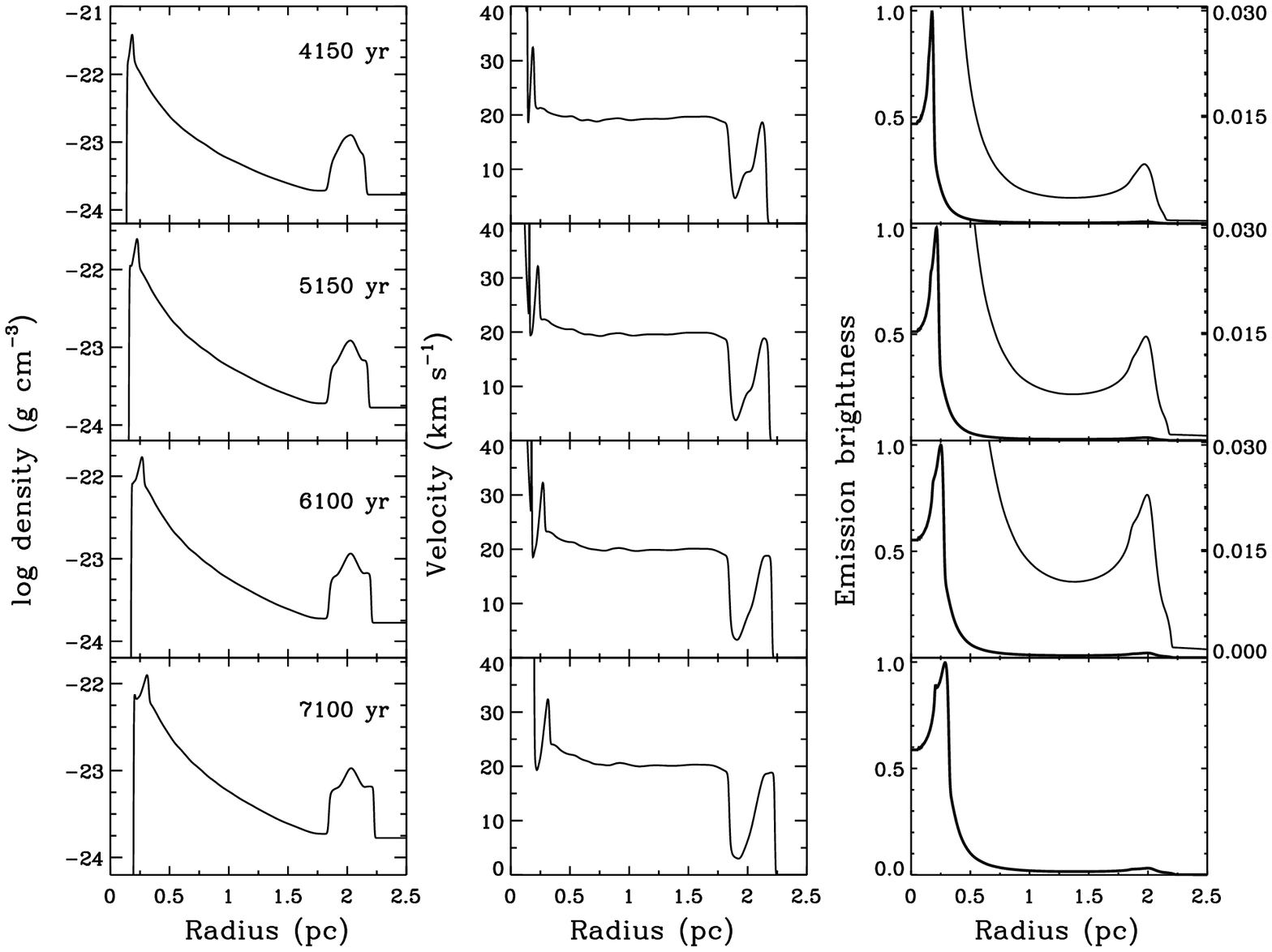}
\caption[ ]{Same as Figure 11, but for the 3.5 \Mso~stellar model.
\label{f15.eps}}
\end{figure}

\begin{figure}
\plotone{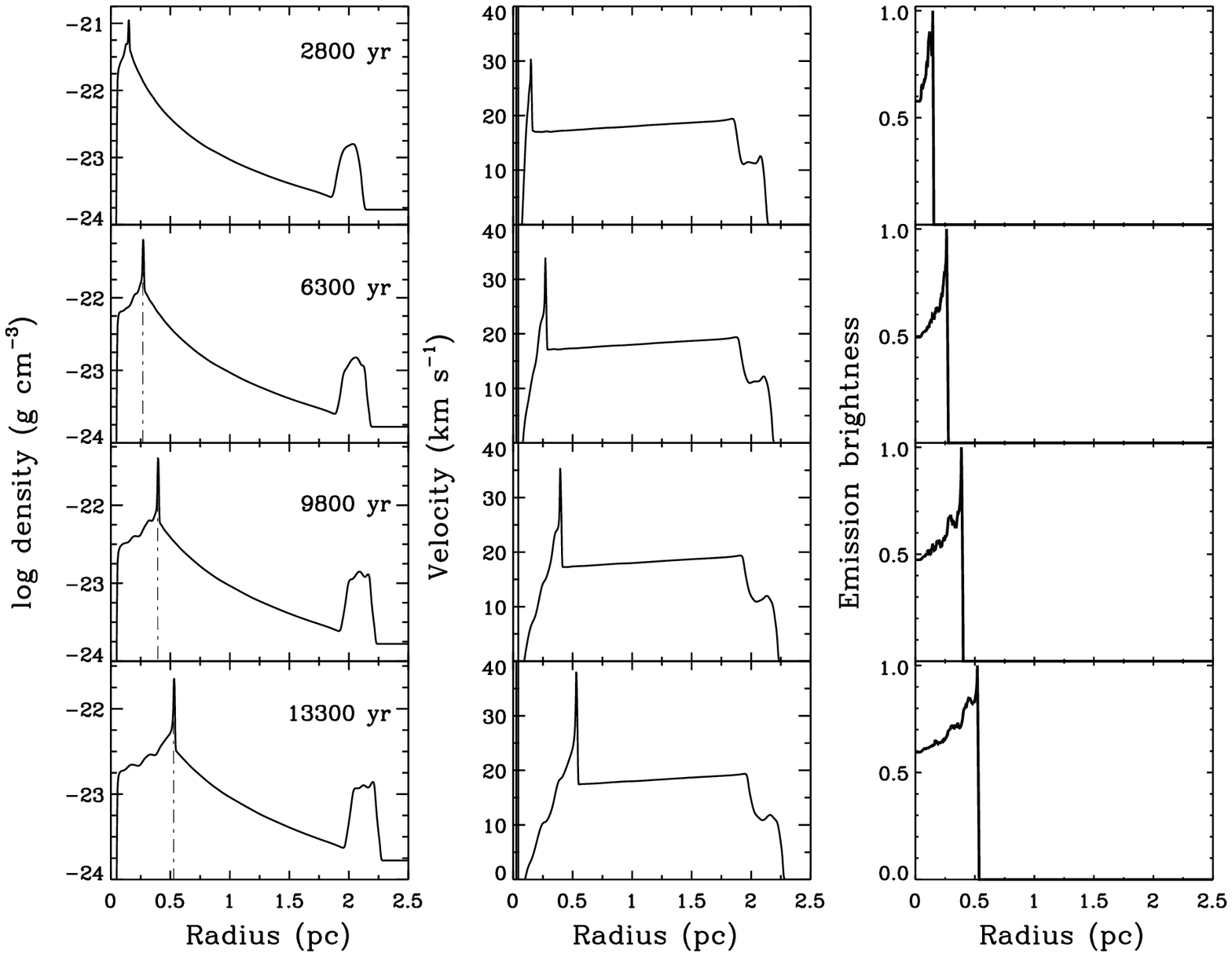}
\caption[ ]{Same as Figure 11, but for the 5 \Mso~stellar model.
\label{f16.eps}}
\end{figure}
 
\begin{figure}
\plotone{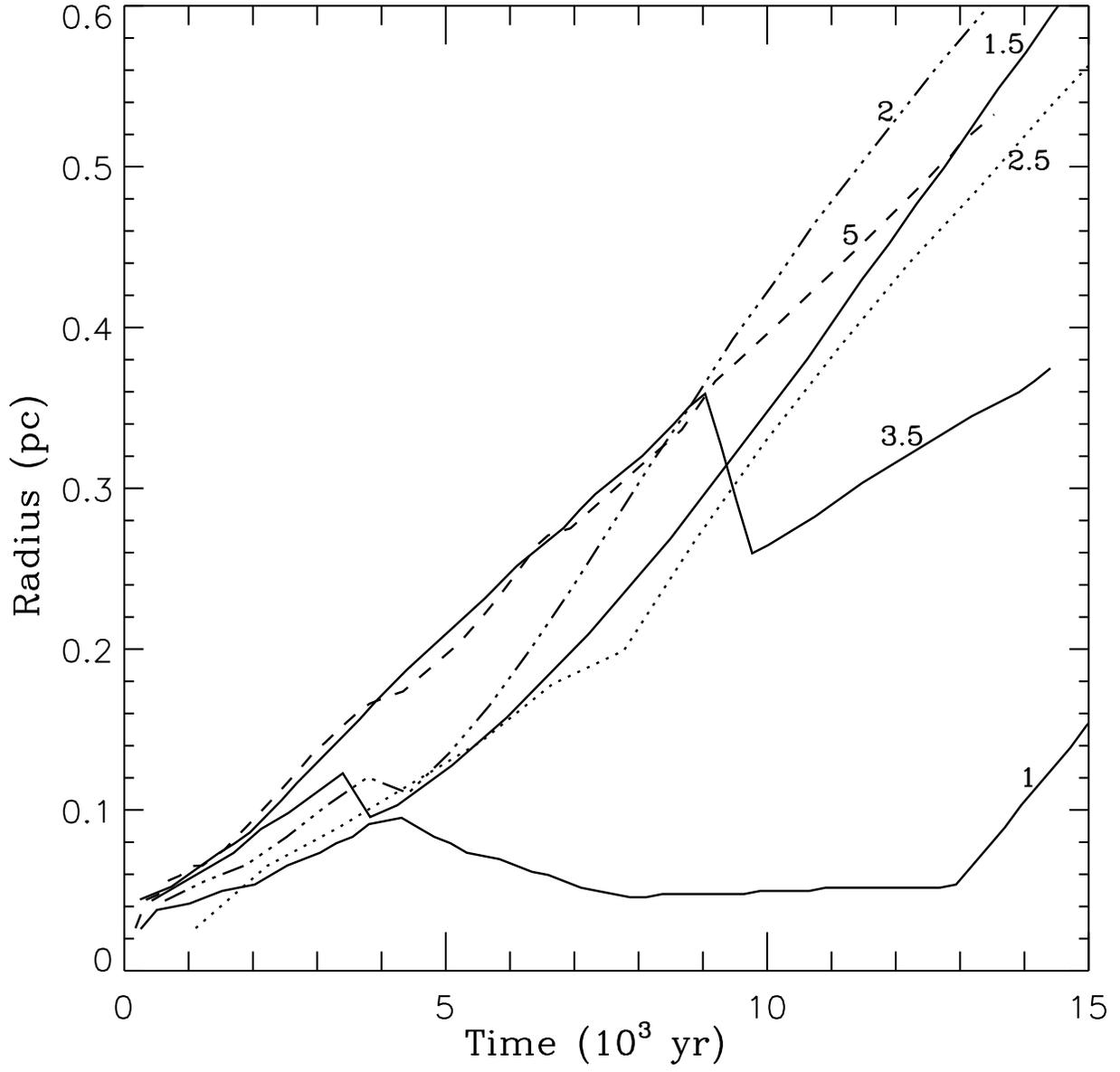}
\caption[ ]{Evolution of the radius of the main shell. The initial mass of
  the model is marked on each line.
\label{f17.eps}}
\end{figure}

\clearpage

\begin{figure}
\epsscale{0.6}
\plotone{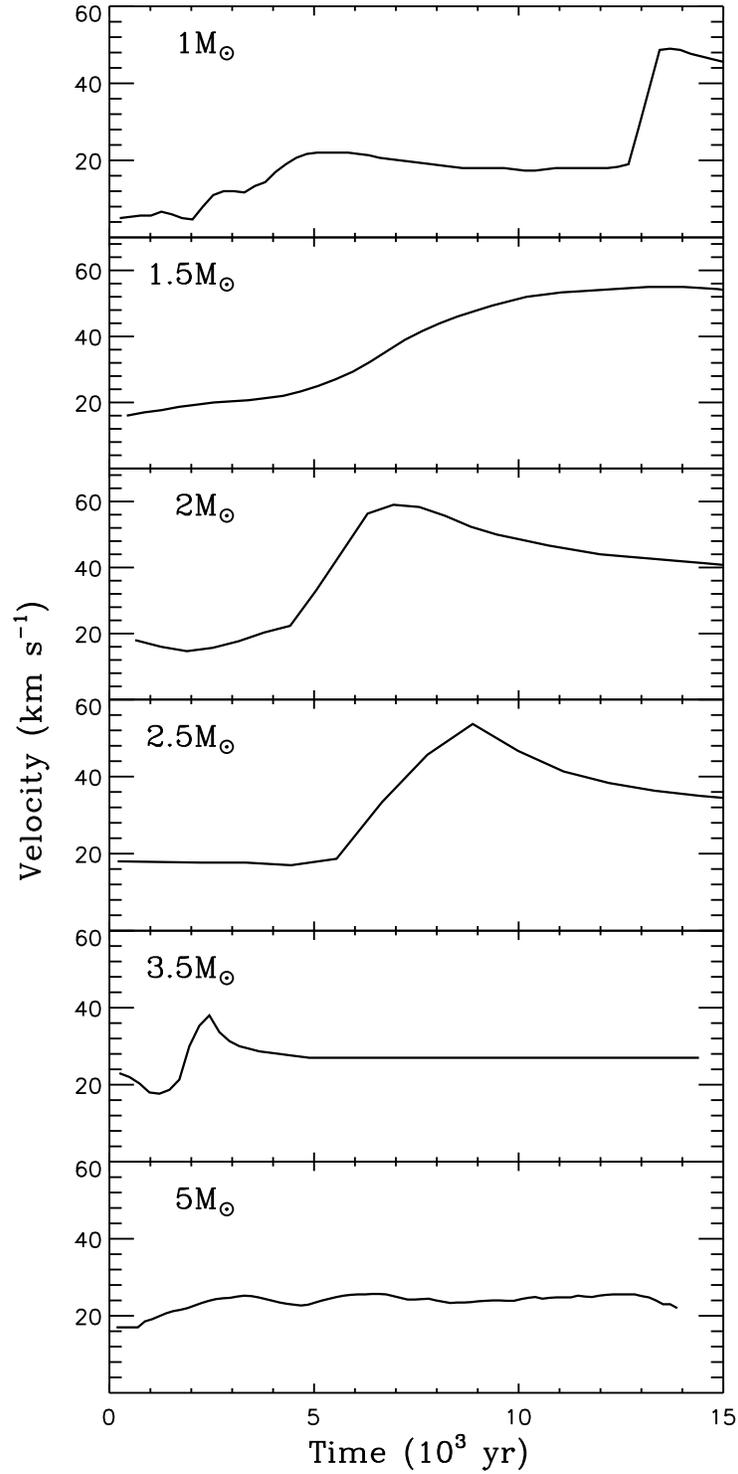}
\caption[ ]{Temporal evolution of the velocity of the main
      shell. From top to bottom, we show the 1, 1.5, 2, 2.5, 3.5 and 5
      \Mso~models. The velocities have been computed by fitting Gaussians to
      the synthetic spectra across the center of the nebula. 
\label{f18.eps}}
\end{figure}

\begin{figure}
\epsscale{1}
\plotone{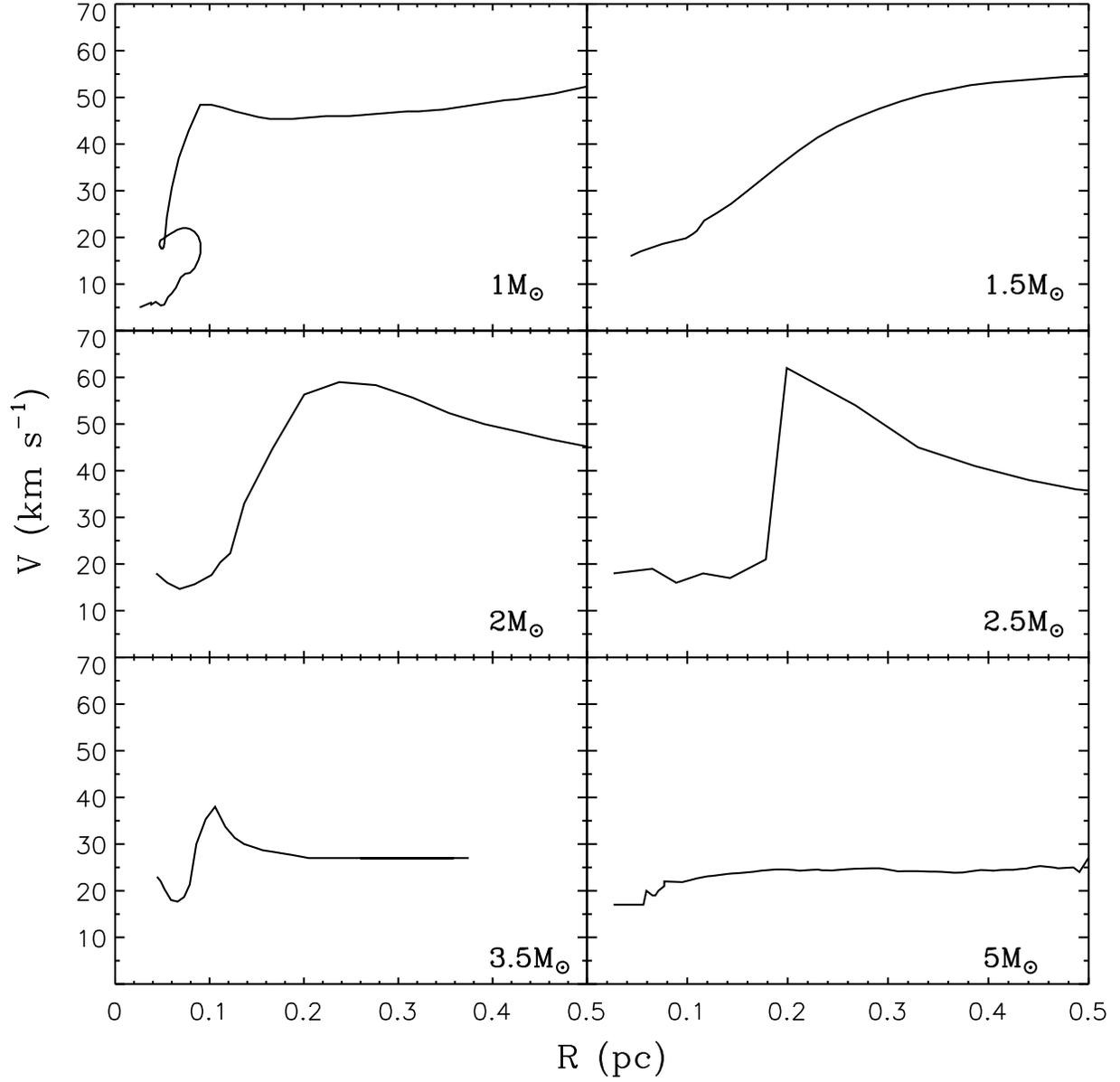}
\caption[ ]{Expansion velocity versus radius for the main shells. From top
  to bottom and from left to right, we show the 1, 1.5, 2, 2.5, 3.5, and 5
      \Mso~models.
\label{f19.eps}}
\end{figure}

\begin{figure}
\plotone{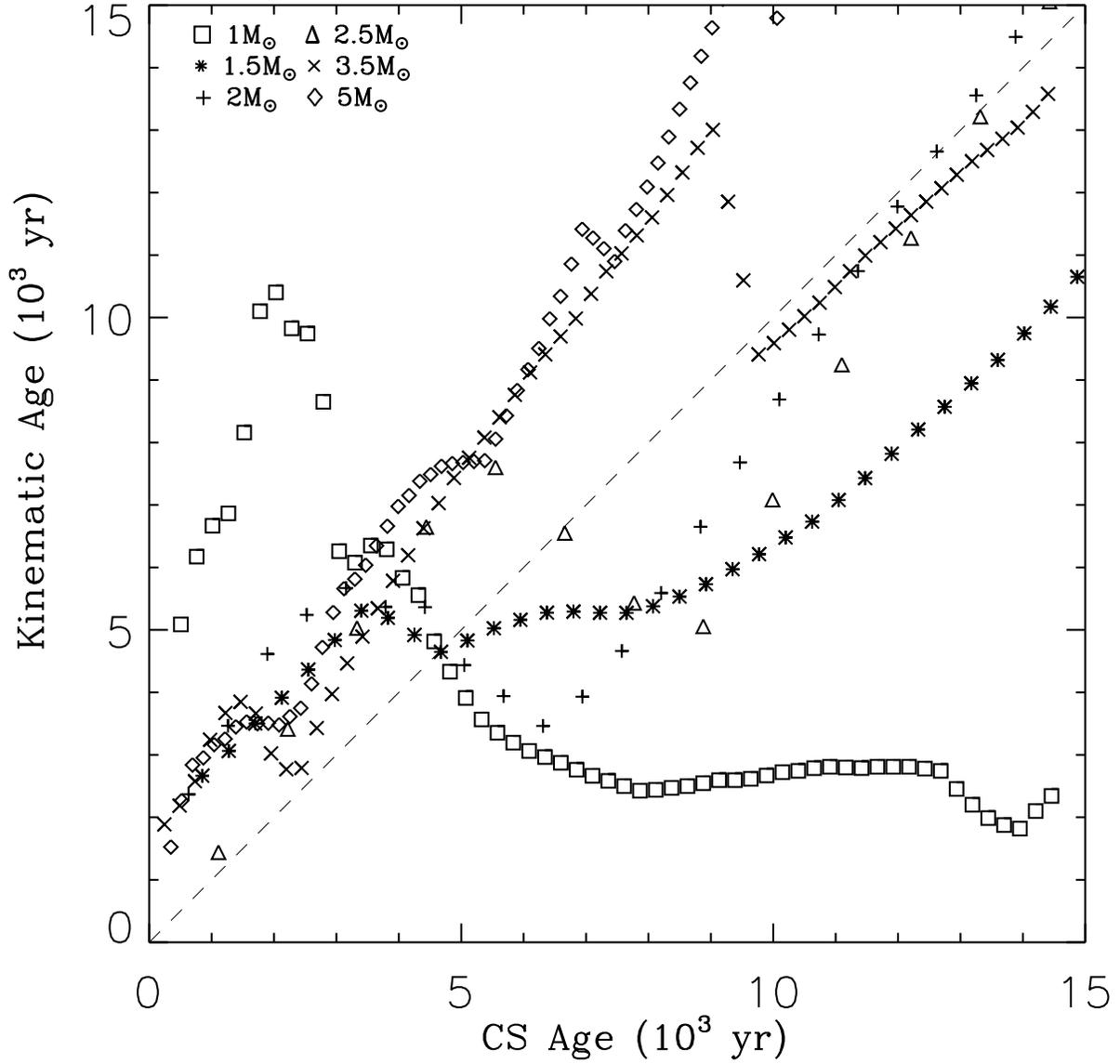}
\caption[]{Kinematical ages derived from our simulations using
      standard procedures are plotted versus the CS ages. The
      symbols 
      used for the different models are shown in the top left corner of the
      plot. 
\label{f20.eps}}
\end{figure}

\begin{figure}
\plotone{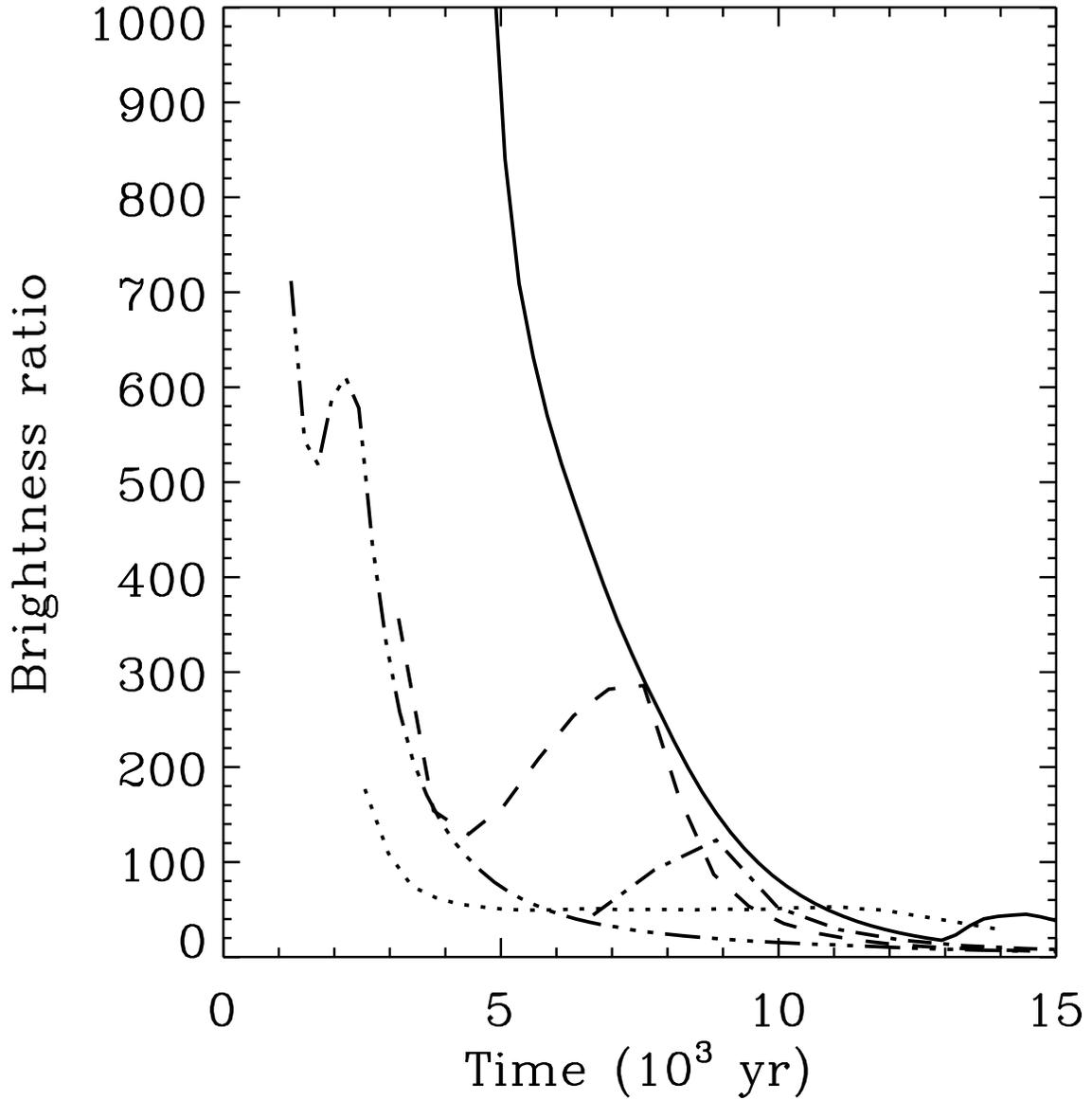}
\caption[ ]{Evolution of the \ha peak emissivity ratio of the inner
      region with respect to that of the detached halos. The solid line
      represents the values for the 1 \Mso~case, while the dotted, dashed,
      dash-dotted, and dash-dot-dot-dotted lines represent the ratio for the
      1, 1.5, 2, 2.5, and 
      3.5 \Mso~models respectively. 
\label{f21.eps}}
\end{figure}

\begin{figure}
\plotone{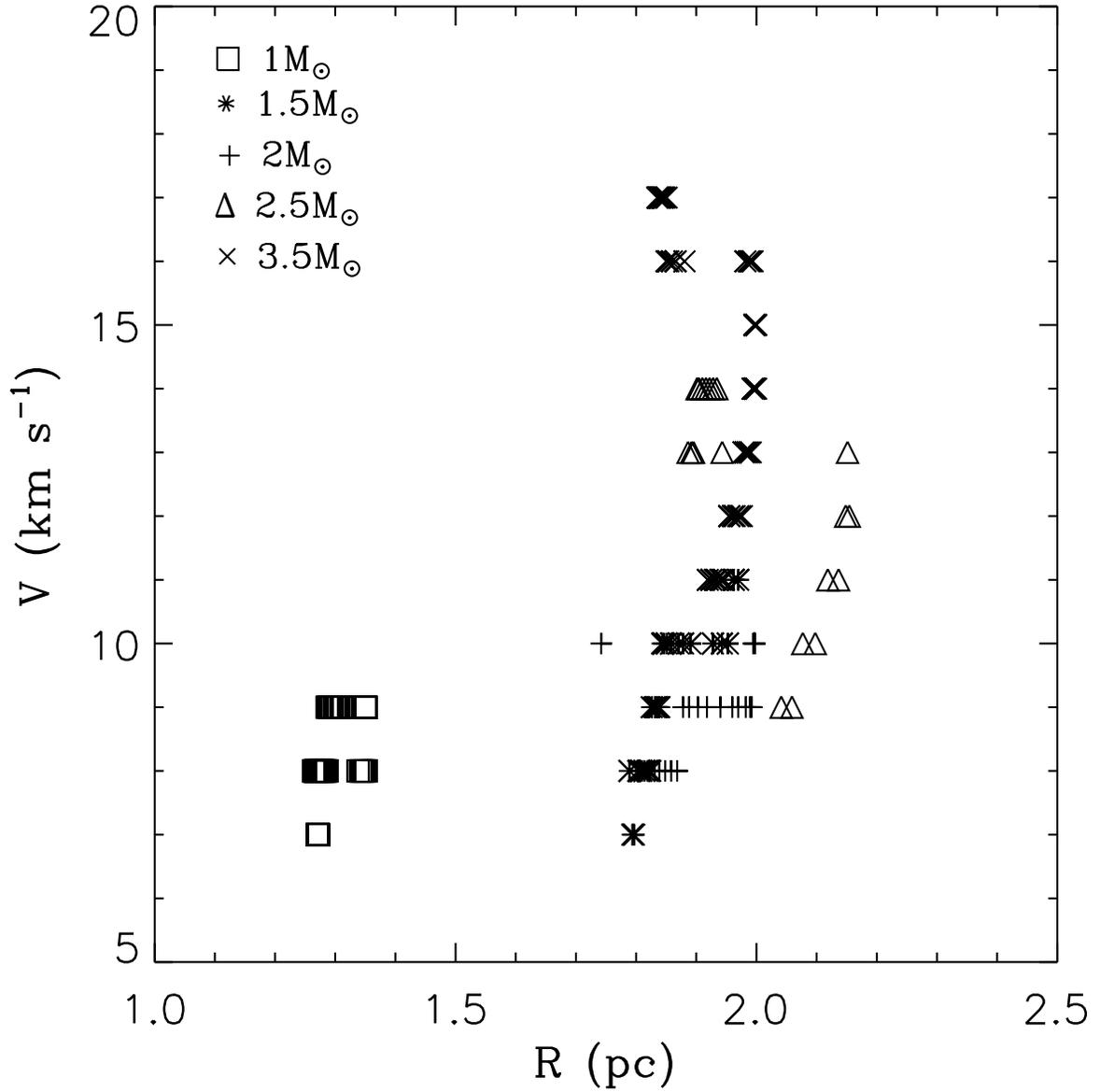}
\figcaption[ ]{Velocity of the outer detached shell
      plotted versus its radius. The temporal evolution of the detached shell
      has been followed up to $10^5$ {\rm yr} and the points represent the
      values measured for each output of the simulations.
\label{f22.eps}}
\end{figure}

\begin{figure}
\plotone{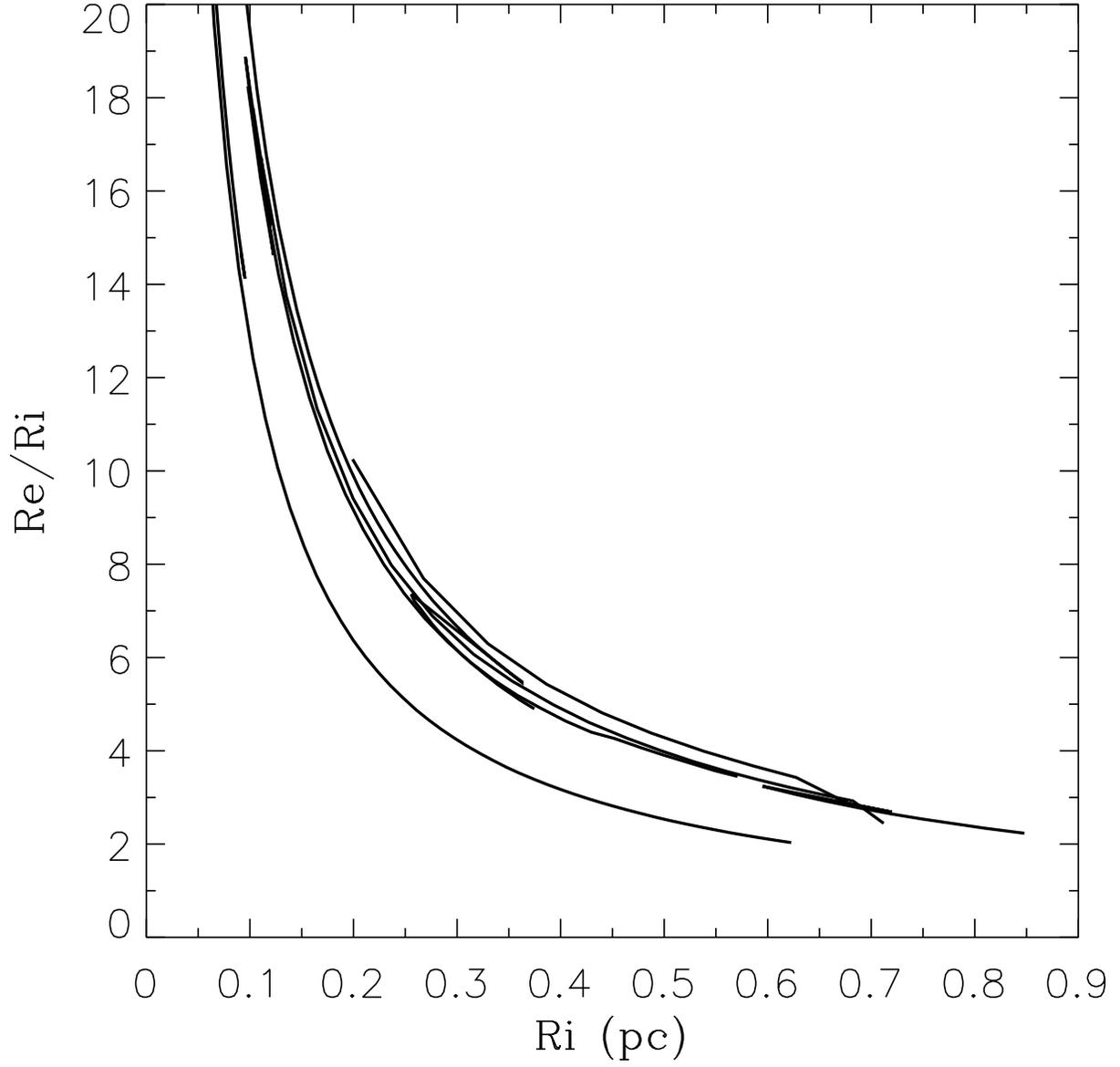}
\caption[ ]{Ratio of the outer
      detached shell to main shell radii ploted versus the radius of the
      main shell. The values shown go from the onset of
      the ionization of the outer shell up to $10^5$ {\rm yr}. 
\label{f23.eps}}
\end{figure}

\begin{figure}
\plotone{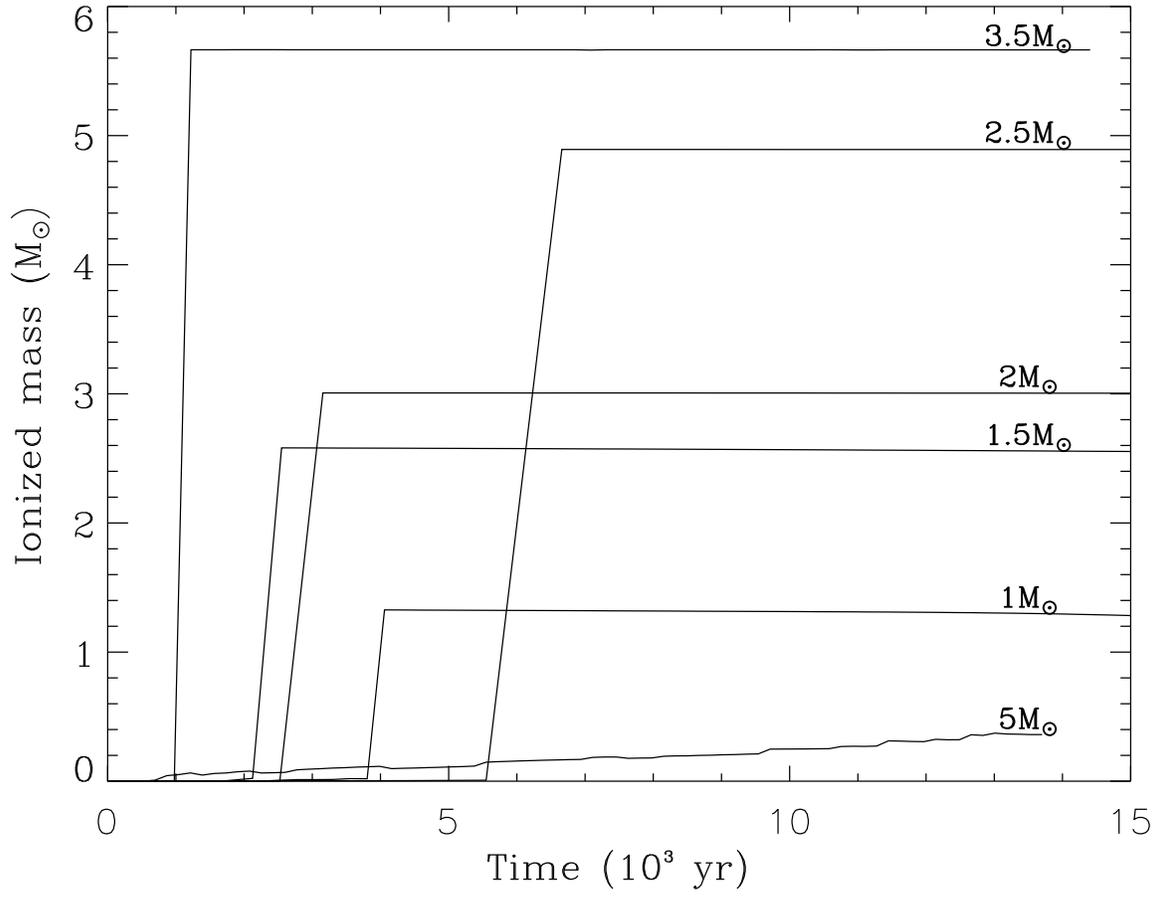}
\caption[ ]{Ionized mass in the grid versus time. The initial mass
      of the model is indicated in each curve.
\label{f24.eps}}
\end{figure}

\begin{figure}
\plotone{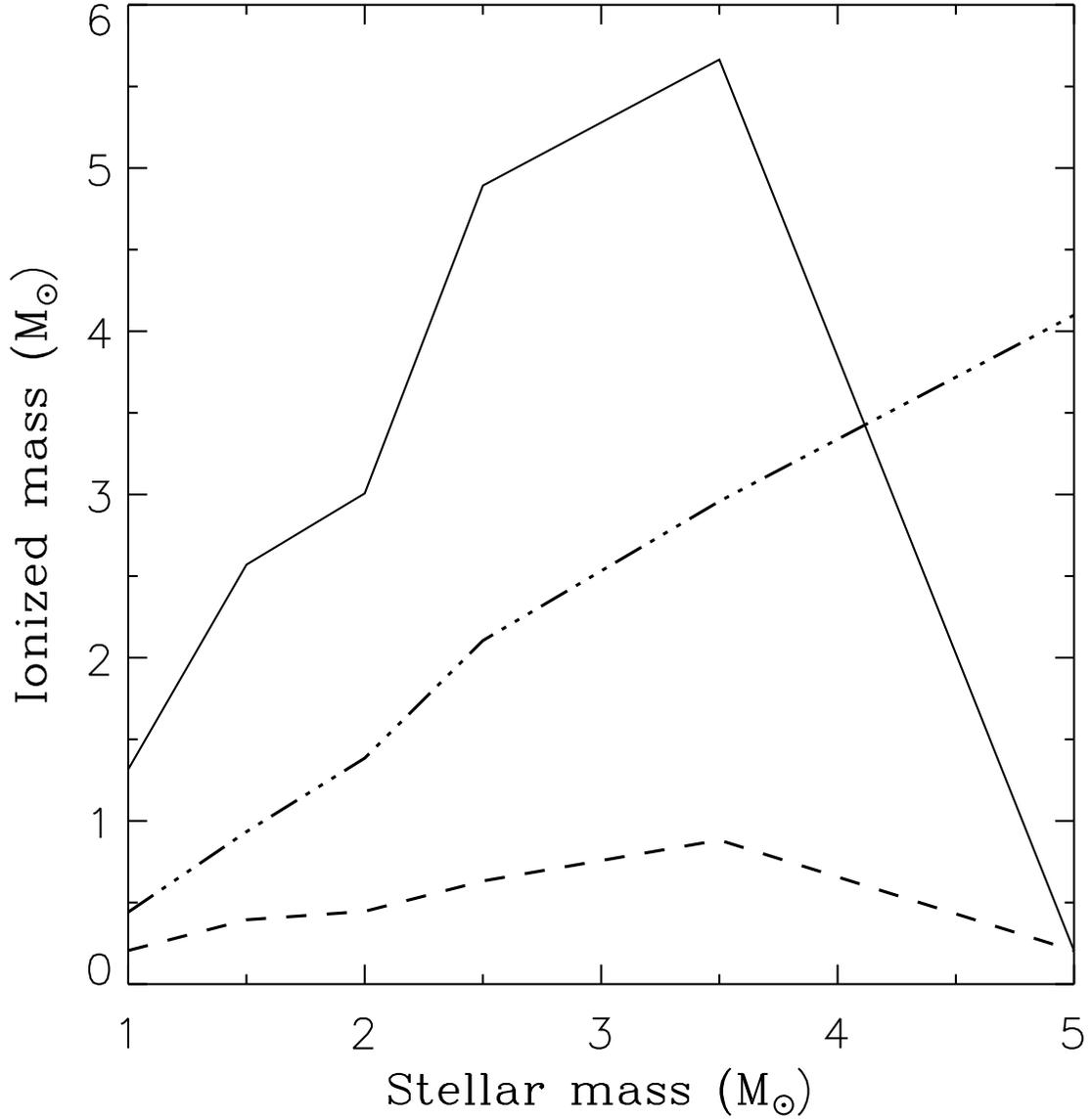}
\caption[ ]{Ionized mass (solid line), total mass lost by the star 
      (dot-dashed line), and ionized mass contained in the main shell
      (dashed line) for the different initial masses considered in the models.
\label{f25.eps}}
\end{figure}


\begin{thebibliography}{}
 
\bibitem[Bianchi(1992)]{Bia:92}
Bianchi, L. 1992, \aap, 260, 314
\bibitem[Balick et al.(1998)]{Betal:98}
Balick, B., Alexander, J., Hajian, A. R., Terzian, Y., 
Perinotto, M. \& Patriarchi, P. 1998, \aj, 116,360
\bibitem[Bl\"ocker(1995)]{Blo:95}
Bl\"ocker, T. 1995, \aap, 297, 727
\bibitem[Bodenheimer, Tenorio-Tagle, \& Yorke(1979)]{Bty:79}
Bodenheimer, G., Tenorio-Tagle, G., \& Yorke, H. W. 1979, \apj, 233, 85
\bibitem[Boffi \& Stanghellini(1994)]{Bs:94}
Boffi, F. R. \& Stanghellini, L. 1994, \aap, 284, 248
\bibitem[Breitschwerdt \& Kahn(1990)]{Bk:90}
Breitschwerdt, D. \& Kahn, F. D. 1990, \mnras , 244, 521
\bibitem[Bryce, Meaburn, \& Walsh(1992a)]{Betal:92a}
Bryce,M., Meaburn, J.,\& Walsh, J.R. 1992, \mnras, 259, 629
\bibitem[Bryce et al.(1992b)]{Betal:92b}
Bryce, M., Meaburn, J., Walsh, J.R., \& Clegg, R.E.S. 1992b, \mnras, 254, 477
\bibitem[Buckley \& Schneider(1995)]{Bs:95}
Buckley, D. \& Schneider, S. E. 1995, \apj,446, 279
\bibitem[Chu, Jacoby, \& Arendt(1987)]{Cja:87}
Chu, Y.-H., Jacoby, G. H., \& Arendt 1987, ApJS, 64, 529.
\bibitem[Chu \& Jacoby(1989)]{Cj:89}
Chu, Y.-H. \& Jacoby, G. H. 1989, in IAU Symposium 131, Planetary Nebulae, 
ed. S. Torres-Peimbert (Dordrecht: Kluwer), 198
\bibitem[Chu et al.(1991)]{Cetal:91}
Chu, Y.-H., Manchado, A., Jacoby, G. H., \& Kwitter, K. B. 1991, \apj, 376,
150 
\bibitem[Corradi et al.(2000)]{Cetal:00}
Corradi, R. L. M., Sch\"onberner, D., Steffen, M., \& Perinotto, M. 2000,
\aap, 354, 1071
\bibitem[Dalgarno \& McCray(1972)]{Dm:72}
Dalgarno, A. \& McCray, R.A. 1972, ARA\&A, 10, 375
\bibitem[Daub(1982)]{Dau:82}
Daub, C. T. 1982, \apj, 612, 624
\bibitem[Dyson et al.(1989)]{Detal:89}
Dyson, J. E., Hartquist, T. W., Pettini, M. \& Smith, L. J. 1989, MNRAS, 241,
625 
\bibitem[Dyson et al.(2000)]{Detal:00}
Dyson, J. E., Hartquist, T. W., Redman, M. P. \&
 Williams, R. J. R. 2000, Ap\&SS, 272, 197
\bibitem[Franco, Tenorio-Tagle, \& Bodenheimer(1989)]{Fttb:89}
Franco, J., Tenorio-Tagle, G., \& Bodenheimer, P. 1989, Rev. Mexicana
Astron. Astrofis., 18, 65
\bibitem[Franco, Tenorio-Tagle, \& Bodenheimer(1990)]{Fttb:90}
Franco, J., Tenorio-Tagle, G., \& Bodenheimer, P. 1990, \apj, 349, 126
\bibitem[Frank, Balick, \& Riley(1990)]{Fbr:90}
Frank, A., Balick, B., \& Riley, J. 1990, \aj, 100, 1903
\bibitem[Frank, van der Veen, \& Balick(1994)]{Fvb:94}
Frank, A., van der Veen, W.E.C.J., \& Balick, B. 1994, \aap, 282, 554
\bibitem[Garc\'{\i}a-Segura \& Franco(1996)]{Gf:96}
Garc\'{\i}a-Segura, G. \& Franco, J. 1996, \apj, 469, 171
\bibitem[Gathier(1984)]{Gath:84}
Gathier, R. 1984, Ph.D thesis, University of Gr\"oningen
\bibitem[Guerrero et al.(1996)]{Getal:96}
Guerrero, M. A., Manchado, A., Stanghellini, L., \& Herrero, A. 1996, 
\apj, 464, 847 
\bibitem[Guerrero, Manchado, \& Chu(1997)]{Gmc:97}
Guerrero, M. A., Manchado, A., \& Chu, Y.-H. 1997, \apj, 487, 328

\bibitem[Guerrero, Villaver, \& Manchado(1998)]{Gvm:98}
Guerrero, M. A., Villaver, E. \& Manchado, A. 1998, \apj, 507, 889

\bibitem[Hajian et al.(1997)]{Hetal:97}
Hajian, A.R., Frank, A., Balick, B., \& Terzian, Y. 1997, \apj, 447, 226.

\bibitem[Kaler(1974)]{Kal:74}
Kaler, J. B. 1974, AJ, 79, 594

\bibitem[Kaler \& Jacoby(1991)]{Kj:91}
Kaler, J. B. \& Jacoby, G. H. 1991, \apj, 372, 215

\bibitem[Kwok, Purton, \& Fitzgerald(1978)]{Kpf:78}
Kwok, S., Purton, C.R., \& Fitzgerald, P.M. 1978, \apj, 219, L125

\bibitem[MacDonald \& Bailey(1981)]{Mb:81}
MacDonald, J. \& Bailey, M. E. 1981, \mnras, 197, 995

\bibitem[Manchado \& Pottasch(1989)]{Mp:89}
Manchado, A. \& Pottasch, S.R. 1989, \aap, 222,226

\bibitem[Manchado et al.(1992)]{Metal:92}
Manchado, A., Guerrero, M. A., Kwitter, K. B., \& Chu, Y.-H. 1992, 
BAAS, 24, 1227

\bibitem[Marten \& Sch\"onberner(1991)]{Ms:91}
Marten, H. \& Sch\"onberner, D. 1991, \aap, 248, 590

\bibitem[McCarthy et al.(1990)]{Mcetal:90}
McCarthy, J. K., Mould, J. R., Mendez, R. H., Kudritzki, R. P.,
Husfeld, D. Herrero, A., \& Groth, H.G. 1990, \apj, 351, 230
\bibitem[Mellema(1994)]{Mel:94}
Mellema, G. 1994, \aap, 290, 915
\bibitem[Middlemass, Clegg, \& Walsh(1989)]{Mcw:89}
Middlemass, D., Clegg, R. E. S., \& Walsh, J. R. 1989, \mnras, 239, 1  

\bibitem[Okorov et al.(1985)]{Oetal:85}
Okorov, V.A., Shustov, B.M., Tutukov, A.V, \& Yorke, H.W. 1985, \aap 142,441


\bibitem[Olofsson et al.(2000)]{Oetal:00}
Olofsson, H., Bergman, P., Lucas, R., Eriksson, K., Gustafsson, B., \&
Bieging, J. H. 2000, A\&A, 353, 583
\bibitem[Osterbrock(1989)]{Ost:89}
Osterbrock, D. E. 1989, Astrophysics of Gaseous Nebulae and Active
Galactic Nuclei (Mill Valley: University Science Books)

\bibitem[Paczy\'nski(1971)]{Pac:71}
Paczy\'nski, B. 1971, AcA, 21, 417, 435 

\bibitem[Pauldrach et al.(1988)]{Petal:88}
Pauldrach, A., Puls, J., Kudritzki, R. H., M\'endez, R. H., \& Heap, S. R.
1988, \aap, 207, 123

\bibitem[Perinotto et al.(1998)]{Petal:98}
Perinotto, M., Kifonidis, K., Sch\"onberner, D., \& Marten , H. 1998, \aap,
332, 1044

\bibitem[Pottasch(1984)]{Pot:84}
Pottasch, S.R., 1984, `Planetary Nebulae', Reidel Publishing Company,
Dordrecht, The Netherlands
\bibitem[Pottasch(1996)]{Pot:96}
Pottasch, S.R., 1996, \aap, 307, 561

\bibitem[Raymond \& Smith(1977)]{Rs:77}
Raymond, J.C. \& Smith, B.W. 1977, ApJS, 35, 419

\bibitem[Sabbadin(1984a)]{Sab:84a}
Sabbadin, F. 1984a, \mnras, 210, 341
\bibitem[Sabbadin(1984b)]{Sab:84b}
Sabbadin, F. 1984b, A\&AS, 58,273

\bibitem[Schmidt-Voigt \& K\"oppen(1987)]{Sk:87}
Schmidt-Voigt, M. \& K\"oppen, J. 1987, \aap, 174, 211

\bibitem[Sch\"onberner et al.(1997)]{Setal:97}
Sch\"onberner, D., Steffen, M., Stahlberg, K., Kifonidis, \& Bl\"ocker, T. 
1997, 
in Advances in Stellar Evolution, ed. R. Wood, \& A. Renzini, 
(Cambridge: Cambridge University Press), 146.


\bibitem[Stanghellini, Corradi, \& Schwarz(1993)]{Scs:93}
Stanghellini, L., Corradi, L.R.M, \& Schwarz, H.E. 1993, \aap, 279, 521

\bibitem[Stanghellini \& Pasquali(1995)]{Sp:95}
Stanghellini, L. \& Pasquali, A. 1995, \apj, 452, 286
\bibitem[ Stanghellini \& Renzini(2000)]{Sr:00}
Stanghellini, L. \& Renzini, A. 2000, \apj, 542, 308
\bibitem[ Stanghellini et al.(2002)]{Setal:02}
Stanghellini, L., Villaver, E., Manchado, A. \& Guerrero, M. A. 2002, \apj,
in press
\bibitem[Stone \& Norman(1992a)]{Sn:92a}
Stone, J.M. \& Norman, M. L. 1992a, ApJS, 80, 753
\bibitem[Stone \& Norman(1992b)]{Sn:92b}
Stone, J.M. \& Norman, M. L. 1992b, ApJS, 80, 791
\bibitem[Stone, Mihalas, \& Norman(1992)]{Smn:92}
Stone, J.M., Mihalas, D., \& Norman, M. L. 1992, ApJS, 80, 819


\bibitem[Vassiliadis \& Wood(1993)]{Vw:93}
Vassiliadis, E. \& Wood, P. 1993, \apj, 413, 641

\bibitem[Vassiliadis \& Wood(1994)]{Vw:94}
Vassiliadis, E. \& Wood, P. 1994, \apj, 92, 125 (VW94)
\bibitem[Villaver, Garc\'{\i}a-Segura, \& Manchado(2002)]{Vgm:02}
Villaver, E., Garc\'{\i}a-Segura, G., \& Manchado, A. 2002, \apj, 571, 880
(Paper I) 

\bibitem[Weinberger(1989)]{Wei:89}
Weinberger, R. 1989, A\&AS, 78, 301

\bibitem[Williams \& Dyson(2002)]{Wd:02}
Williams, R. J. R. \& Dyson, J. E. 2002, MNRAS, 333, 1

\bibitem[Wilson(1948)]{Wil:48}
Wilson, O. C. 1948, \apj, 108,201

\bibitem[Wood \& Faulkner(1986)]{Wf:86}
Wood, P., R. \& Faulkner, D. J. 1986, \apj, 307, 659
\end{thebibliography}
\end{document}